\documentclass[final,11pt,2p,times,sort&compress]{elsarticle} 

  \usepackage{color}
  \usepackage{amsmath}
  \usepackage{calligra}
  \usepackage{caption}
  \usepackage{cprotect}
  \usepackage{enumerate}
  \usepackage[T1]{fontenc}
  \usepackage{ulem}
  \usepackage{lineno}
  \usepackage{multicol}
  \usepackage{tabularx}
  \usepackage[]{titlesec}
  \newcolumntype{L}[1]{>{\raggedright\arraybackslash}p{#1} }
  \newcolumntype{C}[1]{>{\centering  \arraybackslash}p{#1} }
  \newcolumntype{R}[1]{>{\raggedleft \arraybackslash}p{#1} }

\usepackage{mhchem}
\usepackage{longtable} 
\usepackage{pdflscape} 

\journal{Computational Materials Science}

\begin{document}

\begin{frontmatter}
  \title{High-Throughput Prediction of Finite-Temperature Properties using the Quasi-Harmonic Approximation}
  \author{Pinku Nath$^1$, Jose J. Plata$^1$, Demet Usanmaz$^1$, Rabih Al Rahal Al Orabi$^2$, Marco Fornari$^2$, Marco Buongiorno Nardelli$^3$, Cormac Toher$^1$, Stefano Curtarolo$^{7,\star}$}
  \address{$^{1}$ Department of Mechanical Engineering and Materials Science, Duke University, Durham, North Carolina 27708, USA.}
  \address{$^{2}$ Department of Physics and Science of Advanced Materials Program, Central Michigan University, Mount Pleasant, MI 48858, USA.}
  \address{$^{3}$ Department of Physics and Department of Chemistry, University of North Texas, Denton TX, USA.}
  \address{$^{4}$ Materials Science, Electrical Engineering, Physics and Chemistry, Duke University, Durham NC, 27708, USA.}
  \address{$^{\star}${\bf corresponding:} stefano@duke.edu}
  \begin{abstract}
   In order to calculate thermal properties in automatic fashion, the Quasi-Harmonic Approximation (QHA) has been combined with the Automatic Phonon Library (APL) and implemented within the AFLOW framework for high-throughput computational materials science.
   As a benchmark test to address the accuracy of the method and implementation, the specific heats capacities, thermal expansion coefficients, Gr{\"u}neisen parameters and bulk moduli have been calculated for 130 compounds.  
   It is found that QHA-APL can reliably predict such values for several different classes of solids with root mean square relative deviation smaller than 28\% with respect to experimental values. 
   The automation, robustness, accuracy and precision of QHA-APL enable the computation of large material data sets, the implementation of repositories containing thermal properties, 
   and finally can serve the community for data mining and machine learning studies. 
  \end{abstract}
\end{frontmatter}

\section{Introduction}\label{introduction}
The characterization and prediction of thermal properties of materials are amongst the key factors enabling a rational accelerated materials development \cite{nmatHT}. Important properties include specific heat {{capacity}} at constant volume/pressure ($C_V$ or $C_p$),
mode resolved and average  Gr{\"u}neisen parameters ($\gamma_{qj}$ and $\gamma$),
thermal expansion coefficient ($\alpha_V$), Debye temperature ($\theta_D$), lattice thermal conductivity ($\kappa_L$),
and vibrational entropy and Gibbs free energy ($S(p,T)$ and $G(p, T, V)$).

There are several computational techniques leading to the characterization of these thermal properties:
{\bf i.} First principles molecular dynamics (extremely time consuming and computationally impractical for creating large datasets);
{\bf ii.} The {\small GIBBS} approach \cite{BlancoGIBBS2004} also implemented in the {\small AFLOW-AGL}  ({A}utomatic-{G}ibbs-{L}ibrary) \cite{curtarolo:art96}
(very fast and reasonably reliable especially for high-throughput screening \cite{nmatHT});
{\bf iii.} Anharmonic force constant calculations and Boltzmann Transport Equation solvers, as implemented in ShengBTE\cite{ShengBTE_2014}, {\small PHONO3PY} \cite{Togo_prb_2015} and in the {\small AFLOW-APL2} Library \cite{APL2_2016}
(computationally intensive but capable of giving very accurate values for $\kappa_L$);
{\bf iv.} approaches based on the QHA \cite{Taylor1998,Baroni_rmg_2010,Carrier_prb_2007,Alfe_cpc_2009,Duong_jap_2011,Wang_jacers_2014,Huang_cms_2016,Togo_scrmat_2015} which can rapidly characterize $C_V$, $C_p$, $\gamma$, and $\alpha_V$.
Methods {\bf ii-iv.} are based on phonon calculations as available in packages like {\small AFLOW-APL}~\cite{curtarolo:art65,curtarolo:art67}, 
{\small PHONOPY} \cite{Chaput_prb_2011}, Phon \cite{Alfe_cpc_2009}, {\small ALAMODE} \cite{Tadano_jpcm_2014}.

With the goal of creating large repositories of {\it ab-initio} calculated properties, such as in our {\small AFLOW}.org \cite{curtarolo:art75,curtarolo:art92,curtarolo:art104} online database,
we have undertaken the task of implementing the quasi-harmonic method in the {\small AFLOW} software platform \cite{curtarolo:art65}.
The quasi-harmonic method is based on the construction of a strain dependent free energy function in which each strained structure belongs to the
 harmonic regime. 
The strain dependent free energy contributes as a vibrational energy and introduces anharmonic effects into the system, including the temperature dependence.
Although this method has been successfully applied for decades, it has limitations: the QHA loses predictive
power when anharmonic forces play a major role in the dynamics (as in the case of thermal conductivity), under extreme conditions in term of temperature and pressure, \cite{Orlikowski_prb_2006,Xiang_prb2010} or close to their melting point \cite{PhysRevB.79.134106}.
Despite these limitations, this model has been satisfactorily demonstrated to accurately and robustly predict many temperature-dependent properties for compounds of different
nature \cite{Skelton_prb_2014, Iikubo_mtr_2010, Roza_prb_2011, Tohei_mtr_2015, Kangarlou_ijt_2014, burton_jap_2011, Golumbfskie_actamat_2006, Wee_jem_2012}.

Even if the QHA  is a well-established approach, its implementation within an automatic framework requires addressing several challenges.
Therefore, despite the availability of the previously mentioned packages ({\small PHONOPY}, and {\small ALAMODE}), to the best of our knowledge,
there is not yet a high-throughput \cite{nmatHT} framework able to predict temperature dependent properties using  the QHA in  a self-contained robust way.
The high-throughput protocol should include: automatic generation of files, robust correction of errors and post-processing, and appropriate interface to a material database \cite{curtarolo:art75}.
In this article, we show tests of our QHA implementation in {\small AFLOW} by computing
temperature dependent thermodynamic properties for more than one hundred materials. For one case we assess the effect of improved electronic structure \cite{curtarolo:art93} on the thermal properties.

\section{Methods\label{methods}}

\subsection{ Ab initio thermodynamics}

In the framework of the QHA, the Helmholtz free energy, $F$, for a fixed number of particles, is written as
\begin{equation}\label{EQ:GIBBS}
 F(V, T)= E_{0K}(V)+ F_{vib}(V,T) + F_{elec}(V,T)
\end{equation}
where $E_{0K}$ is the total energy of the system at 0K and given volume, $V$.  $F_{vib}$ represents the vibrational contribution to the free energy and $F_{elec}$ is the
electronic contribution to the free energy as function of volume and temperature.
The total energy at any volume and 0 K can be computed by using standard periodic quantum mechanical software such as Quantum Espresso~\cite{quantum_espresso_2009} or the Vienna Ab-initio Simulation Package (VASP)~\cite{kresse_vasp} and properly relaxed structures. The vibrational free energy, which includes zero point energy contributions,  can be obtained from the phonon density of states, $g(\nu)$, via:
\begin{equation}\label{EQ:FVIB}
  F_{vib}(V, T)= \int_{0}^{\infty}g(\nu)\bigg[\frac{h\nu}{2}+k_B T \text{ln}\Big(1-\text{exp}\big(-\frac{h\nu}{k_BT}\big)\Big)\bigg]\text d\nu
\end{equation}
where $k_B$, $h$, and $\nu$ are the Boltzmann constant, the Planck constant, and the vibrational frequency respectively.
Frequencies for a given wave vector $\mathbf{q}$ can be computed by diagonalizing the dynamical matrix. The phonon density of states, pDOS, can be computed by integrating the phonon dispersion in the Brillouin zone.

Similarly, $F_{elec}$, can be computed as:
\begin{equation}\label{EQ:FELE}
 F_{elec}(V,T)=\Delta E_{elec}(V,T)-\text TS_{elec}(V,T)
\end{equation}
where $\Delta E_{elec}(V,T)$ and $S_{elec}(V,T)$ are the contribution to the electronic energy due to temperature changes and the electronic entropic contribution to the free energy.
At  low  temperatures, $\Delta E_{elec}(V,T)$ is  very  small  and  can be neglected. However, it may play a significant role at high temperatures.
Both can be computed using the electronic density of states, eDOS,
\begin{equation}\label{EQ:EELE}
\Delta E_{elec}(V,T)=\int n(\epsilon) f \epsilon d\epsilon - \int^{\epsilon_F} n(\epsilon)\epsilon \text d\epsilon
\end{equation}
\begin{equation}\label{EQ:SELE}
S_{elec}(V,T)=-k_B\int n(\epsilon)[f \text{ln} f + (1-f)\text{ln}(1-f)]\text{d}\epsilon
\end{equation}
where the eDOS at energy, $\epsilon$, is represented by $n(\epsilon)$, and $f$ is the Fermi distribution function.

  Once $F(V, T)$ is computed at different volumes and temperatures, extracting the thermodynamic data is a straightforward process using the equations of state.
  For instance, properties like equilibrium free energy, $F_{eq}$, equilibrium volume, $V_{eq}$, bulk modulus, $B$, and  the derivative of the bulk modulus with respect to pressure, $B_p$, can be obtained by fitting $F(V, T)$  at different volumes and temperatures to the Birch-Murnaghan (BM) function:
\begin{equation}\label{EQ:BM}
F(V) = F_{eq} + \frac{BV_{eq}}{B_p}\Bigg[ \frac{(V_{eq}/V)^{B_p}}{Bp-1} +1\Bigg] - \frac{V_{eq}B}{Bp-1} 
\end{equation}
 where, $F_{eq}$, $B$, $V_{eq}$ and $B_p$ are used as the fitting parameters.

\par
  The mode Gr{\"u}neisen parameters, $\gamma_{qj}$, for the wave vector $\mathbf{q}$ and the phonon branch $j$ can be computed by taking the derivative of the  dynamical matrix with respect to the volume, as \cite{ThermoCrys}:
\begin{equation}\label{EQ:GAMMAQJD}
 \gamma_{qj}=-\frac{V_{eq}^{0K}}{2\nu_{qj}^2}\sum_j e_{qj} \frac{\partial D_q}{\partial V} e_{qj}^*
 \end{equation}
where $D_q$ is the dynamical matrix for a wave-vector, $\mathbf{q}$, $\nu_{qj}$ vibrational frequency,  and $e_{qj}$ is the eigenvector for phonon branch, $j$. An average Gr{\"u}neisen parameter, $\gamma(T)$, can be obtained using \cite{PhysPhon,IntroLattDym}:
\begin{equation}\label{EQ:GAMMA}
 \gamma(T)=\frac{\sum_{q,j}\gamma_{qj}C_{V_{qj}}}{C_V}
 \end{equation}
 where $C_{V_{qj}}$, is the isochoric specific heat:
 \begin{equation}\label{CV}
 C_{V_{q,j}}=k_B\sum_{q,j} \frac{(h\nu^2_{qj})\exp{(h\nu_{qj}/k_BT)}}{(k_BT)^2(\exp(h\nu_{qj}/k_BT)-1)^2}
 \end{equation}
   Once $\gamma(T)$ is calculated, other variables such as volumetric thermal expansion $\alpha_V(T)$ and isobaric specific heat, $C_p$ can be predicted using Eq. \ref{EQ:ALPHA} and Eq. \ref{EQ:CP} respectively:
 \begin{equation}\label{EQ:ALPHA}
\alpha_V(T)=\frac{C_V(T)\gamma(T)}{V(T)B(T)}
\end{equation}
\begin{equation}\label{EQ:CP}
C_p-C_V=\alpha^2(T)B(T)V(T)T
\end{equation}

\subsection{Computational details}

In the QHA-APL we first perform a geometry optimization minimizing the forces acting on the atoms in the primitive cell  and the stresses.  The optimized geometry is used as  starting point for the other calculations. The phonon dispersions are computed at three different volumes to determine the Gr{\"u}neisen parameters, one at the equilibrium volume and the other two at slightly distorted volumes (less than $\pm 5\%$ of the volume). Finally, the data are used to fit the BM equation of state.  These calculations are automatically generated, managed and monitored by the {\small AFLOW} \cite{curtarolo:art65,curtarolo:art92} package, facilitating and accelerating the prediction of all properties required by the user in the original input.

\subsubsection{Geometry optimization}

All structures are fully relaxed  using the HT framework, {\small AFLOW} \cite{curtarolo:art65},
and the DFT Vienna Ab-initio simulation package, {\small VASP} \cite{kresse_vasp}. Optimizations are performed following the
{\small AFLOW} standards \cite{curtarolo:art104}. We use the projector augmented wave (PAW) pseudopotentials \cite{PAW}
and the exchange and correlation functionals parametrized by the generalized gradient approximation proposed
by Perdew-Burke-Ernzerhof (PBE) \cite{PBE}. All calculations use a high energy-cutoff, which is 40$\%$ larger
than the maximum recommended cutoff among all component potentials, and a k-points mesh of 8000 k-points per reciprocal atom.
Primitive cells are fully relaxed (lattice parameters and ionic positions) until the energy difference between two consecutive ionic steps
is smaller than $10^{-4}$ eV and forces in each atom are below $10^{-3}$ eV/\AA.

\subsubsection{Phonon calculations}

Phonon calculations  were  carried out  using the automatic phonon library, APL, as implemented in the {\small AFLOW} package, using  VASP  to obtain the interatomic force constants (IFCs) via the finite-displacement approach.
The magnitude of this displacement is 0.015 \AA. 
Non-analytical contributions to the dynamical matrix are also included using the formulation developed by Wang \textit{et al} \cite{Wang2010}.
Frequencies and other related phonon properties are calculated on a  21$\times$21$\times$21 mesh in the Brillouin zone, which is sufficient to converge the  vibrational  density of  states, pDOS,  and  hence  the  values  of thermodynamic properties calculated through it.
 The pDOS is calculated using the linear interpolation tetrahedron method available in {\small AFLOW} package.
The derivative of dynamical matrix in Eq. \ref{EQ:GAMMAQJD} is obtained using the central difference method within a volume range of $\pm 0.03 \%$.

\par
\begin{figure}[h!]
  \centerline{\includegraphics*[width=0.70\textwidth,clip=true]{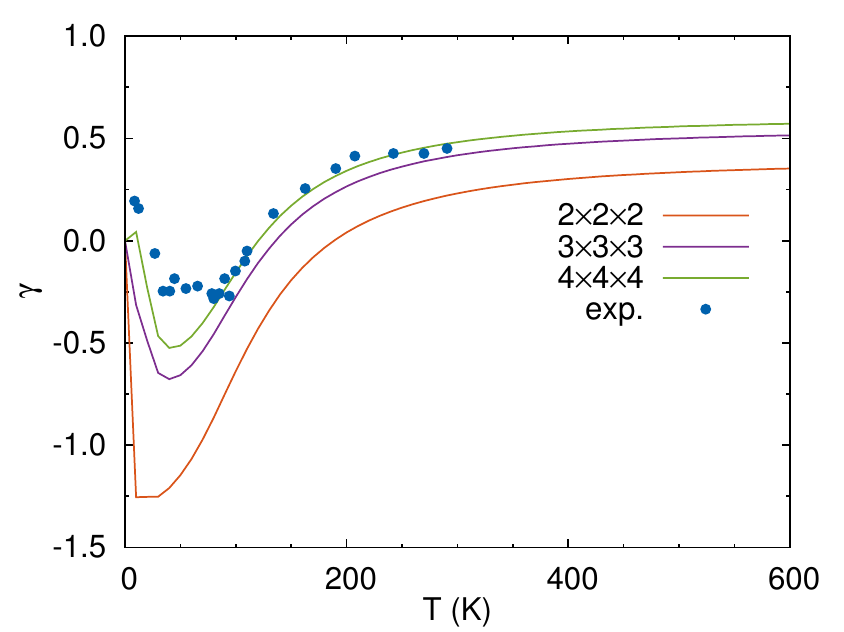}}
\vspace{-1mm}
  \caption{\small Average Gr{\"u}neisen parameter for three different super-cell sizes for Si. Circles represent experimental values \cite{Gauster_prb_1971}.}
  \label{fig:SUPERCELL}
\end{figure}

 In order to balance accuracy and computational cost, the dimension
of the supercell for IFCs calculations has been optimized. Since the average Gr{\"u}neisen parameter is an important variable in our equations, we have evaluated this property for different cell size for Si. Results and comparison with experimental results are shown in Fig. \ref{fig:SUPERCELL}. We note that a $3\times 3\times 3$ supercell, containing 54 atoms,  predicts almost the same quantitative values as the $4\times 4\times 4$ supercell (128 atoms) in the 100-700 K range, while dramatically reducing  the number of atoms and computational time. Therefore, we have built supercells with at least 27 atoms per reciprocal atom in the primitive cell homogeneously distributed in all directions.

\subsubsection{Distorted volume single point calculations}
 To obtain $F(V,T)$ vs. $V(T)$ curves, phonon calculations have been performed for thirteen equally spaced configurations in which the volume of cell was compressed and expanded in the range from 98 $\%$ to 113 $\%$ of its fully relaxed value. $F_{elec}(V,T)$ is computed from static and band calculations of the expanded and compressed primitive cells following {\small AFLOW} standard \cite{curtarolo:art104} that are consistent with the setup used for the geometry optimizations.

There are limitations derived by the level of theory or functional.
  PBE could predict metallic behavior in semiconductors with a narrow band gap.
  This error  would  modify $F_{elec}(V,T)$ which is a small contribution at low and moderate temperatures. To verify that none of the materials included in the data set present this anomaly, we have compared the calculated band gaps with the experimental values (see Table 1 in  Supplementary Information).

{{
\subsection{Analysis of Results}

Different statistical parameters have been used to measure the qualitative and quantitative agreement of QHA-APL with experimental values. 
The Pearson correlation coefficient $r$ is a measure of the linear
correlation between two variables, $X$ and $Y$. It is calculated by
\begin{equation} 
\label{Pearson}
 r = \frac{\sum_{i=1}^{n} \left(X_i - \overline{X} \right) \left(Y_i - \overline{Y} \right) }{ \sqrt{\sum_{i=1}^{n} \left(X_i - \overline{X} \right)^2} \sqrt{\sum_{i=1}^{n} \left(Y_i - \overline{Y} \right)^2}},
\end{equation}
where $\overline{X}$ and $\overline{Y}$ are the mean values of $X$ and $Y$.

The Spearman rank correlation coefficient $\rho$ is a measure of the monotonicity of the relationship between two variables.
The values of the two variables $X_i$ and $Y_i$ are sorted in ascending order, and are assigned rank values $x_i$ and $y_i$ which
are equal to their position in the sorted list. The correlation coefficient is then given by
\begin{equation} 
\label{Spearman}
 \rho = \frac{\sum_{i=1}^{n} \left(x_i - \overline{x} \right) \left(y_i - \overline{y} \right) }{ \sqrt{\sum_{i=1}^{n} \left(x_i - \overline{x} \right)^2} \sqrt{\sum_{i=1}^{n} \left(y_i - \overline{y} \right)^2}}.
\end{equation}
It is useful for determining how well the ranking order of the values of one variable predict the ranking order of the values of the other variable.

We also investigate the root-mean-square relative deviation ($RMSrD$)
of the calculated results from experiment. This gives a measure of the magnitude of the difference between the QHA-APL predictions and experiment. The
root mean square relative deviation ($RMSrD$) is calculated using the expression
\begin{equation} 
\label{RMSD}
RMSrD  = \sqrt{\frac{ \sum_{i=1}^{n} \left( \frac{X_i - Y_i}{X_i} \right)^2 }{N - 1}} ,
\end{equation}
Note that lower values of the $RMSrD$ indicate better agreement with experiment.
}}

\section{Results}\label{results}

A benchmark of 130 materials has been used to demonstrate the accuracy and robustness of this method (see Table~\ref{tab:data}).
To maximize the heterogeneity of the data set, these compounds were selected to belong to different crystallographic
lattices (cubic, tetragonal, orthorhombic, hexagonal, rhombohedral and monoclinic) as well as different electronic properties (insulators, semiconductors and metals, see Supplementary Information for the computed and experimental energy gaps).

\begin{landscape}
\begin{scriptsize}
\begin{longtable}[h]{c r l l l l | c r l l l l}
  \caption{{{\small ICSD identification number, Bulk modulus ($B$), Gr{\"u}neisen parameter ($\gamma$), Volumetric thermal expansion coefficient ($\alpha$), and Isochoric specific heat ($C_p$) for material data set. Experimental values in parenthesis. Units: $B$ in GPa, $\alpha$ in 10$^{-6}$K$^{-1}$, and $C_p$ in Jmol$^{-1}$K$^{-1}$ (see Supplementary Information for the computed and experimental energy gaps).}}}\\
\hline  Formula           &  ICSD    &    $B$ & $\gamma$ & $\alpha$ & C$_p$ &  Formula         &  ICSD      & $B$      & $\gamma$ &  $\alpha$    & C$_p$ \\ \hline
\endfirsthead
\hline  Formula           &  ICSD    &    $B$ & $\gamma$ & $\alpha$ & C$_p$ &  Formula         &  ICSD      & $B$      & $\gamma$ &  $\alpha$    & C$_p$ \\ \hline
\endhead
\hline
\endfoot
\hline \hline
\endlastfoot
  \ce{Ag1Cl1}      & 157535   &   28.6(44.0)\cite{Hughes_prb_1996}                     & 2.30                                           &   114.0                                   &  53.5(53.0)\cite{IhsanBarin}         &
  \ce{Ge1}         &  44841   &   55.1(78.0)\cite{Madelung_Semiconductors_2004}        & 0.74(0.76)\cite{slack}                         &    21.5(16.2)\cite{Lide_CRC_2004}         &  23.3(23.3)\cite{IhsanBarin} \\
  \ce{Ag1Mg1}      & 184205   &   48.6                                                 & 2.14                                           &    89.2                                   &  50.5                                &
  \ce{Ge1Mg2}      &  81735   &   46.63                                                & 1.46                                           &    53.8                                   &  71.2 \\
  \ce{Ag1Sc1}      & 58348    &   65.8                                                 & 1.66                                           &    46.6                                   &  49.3                                &
  \ce{H1Li1}       &  61751   &   32.0(33.7)\cite{Laplaze_ssc_1976}                    & 1.10(1.28)\cite{slack}                         &    87.1                                   &  27.6(28.1)\cite{IhsanBarin} \\
  \ce{Ag3Mg1}      & 58323    &   58.1                                                 & 2.41                                           &    86.9                                   & 102.7                                &
  \ce{H1Li1Pd1}    &  246613  &   77.5                                                 & 1.71                                           &    63.4                                   &  17.7 \\
  \ce{Al1As1}      & 606008   &   64.5(77.0)\cite{Lam_prb_1987}                        & 0.57(0.66)\cite{slack}                         &    13.7                                   &  44.4(45.8)\cite{IhsanBarin}         &
  \ce{H1Mg1Ni1}    &  187257  &   86.3                                                 & 1.21                                           &    40.1                                   &  55.1 \\
  \ce{Al1B2}       & 159334   &  167.5                                                 & 1.16                                           &    22.2                                   &  51.0(43.9)                          &
  \ce{H1Na1}       &  183291  &   18.8                                                 & 0.79                                           &    78.9                                   &  34.8(36.5) \\
  \ce{Al1}         & 240129   &   67.5                                                 & 2.41                                           &    80.8(69.0)\cite{Lide_CRC_2004}         &  24.3(24.2)\cite{Lide_CRC_2004}      &
  \ce{H1Ti1}       &  168325  &  112.8                                                 & 0.88                                           &    15.8                                   &  26.4 \\
  \ce{Al1Li1}      & 240121   &   47.6                                                 & 1.54                                           &    69.1                                   &  44.5(48.9)\cite{IhsanBarin}         &
  \ce{Hg1Ni1}      &  639119  &  104.8                                                 & 2.82                                           &    68.2                                   &  51.3 \\
  \ce{Al1N1}       & 602459   &  188.0(201.0)\cite{McNeil_jacers_1993}                 & 0.80(0.70)\cite{Morelli_Slack_2006}            &    10.1                                   &  31.0(30.1)\cite{IhsanBarin}         &
  \ce{Hg1Pd1}      &  639137  &  101.0                                                 & 3.10                                           &    67.2                                   &  52.0 \\
  \ce{Al1Ni1}      & 608805   &  147.1                                                 & 1.93                                           &    40.5                                   &  45.9(45.9)\cite{IhsanBarin}         &
  \ce{Hg1Pt1}      &  104337  &  127.4                                                 & 3.28                                           &    57.0                                   &  51.8 \\
  \ce{Al1P1}       & 609019   &   80.6(86.0)\cite{Lam_prb_1987}                        & 0.50(0.75)\cite{Morelli_Slack_2006}            &    10.3                                   &  41.5(42.0)\cite{IhsanBarin}         &
  \ce{Hg1Zr1}      &  639318  &   97.6                                                 & 2.19                                           &    41.3                                   &  49.9 \\
  \ce{Al1Sb1}      & 609290   &   47.3(58.2)\cite{Madelung_Semiconductors_2004}        & 0.44(0.60)\cite{slack}                         &    11.7(12.6)\cite{Lide_CRC_2004}         &  45.8                                &
  \ce{I1K1}        &  53827   &    8.3(11.1)\cite{Hauss_zfp_1960}                      & 1.77(1.45)\cite{slack}                         &   178.6(122.4)\cite{Li_jpcrd_1976}        &  54.0(52.80)\cite{Li_jpcrd_1976} \\
  \ce{Al1Sc1}      & 58098    &   63.6                                                 & 1.83                                           &    52.9                                   &  47.0                                &
  \ce{I1Li1}       &  27983   &    9.7                                                 & 2.93                                           &   395.9(178.2)\cite{Li_jpcrd_1976}        &  65.1 \\
  \ce{Al1Si1Sr1}   & 162865   &   51.9                                                 & 1.31                                           &    37.7                                   &  69.3                                &
  \ce{I1Na1}       &  52240   &   14.0(15.95)\cite{Hauss_zfp_1960}                     & 1.94(1.56)\cite{slack}                         &   156.0(136.5)\cite{Li_jpcrd_1976}        &  53.5(52.26)\cite{Li_jpcrd_1976} \\
  \ce{Al1Tb3}      & 58173    &   45.9                                                 & 0.51                                           &    16.2                                   &  96.5                                &
  \ce{I1Rb1}       &  53846   &    7.1(11.1)\cite{Hauss_zfp_1960}                      & 1.54(1.41)\cite{slack}                         &   161.6(124.5)\cite{Li_jpcrd_1976}        &  53.3(52.5)\cite{IhsanBarin}\\
  \ce{Al1Ti1}      & 187030   &   92.1                                                 & 1.54                                           &    35.6                                   &  45.6(49.3)\cite{IhsanBarin}         &
  \ce{In1N1}       &  157515  &  118.5(126.0)\cite{Ueno_prb_1994}                      & 0.82(0.97)\cite{Krukowski_jphyschemsolids_1998}&    14.3                                   &  40.0 \\
  \ce{Al3Ti1}      & 609525   &    97.9                                                & 1.49                                           &    33.8                                   &  89.7                                &
  \ce{In1P1}       &  165466  &   56.5(71.0)\cite{Lam_prb_1987}                        & 0.62(0.60)\cite{slack}                         &    15.7(13.8)\cite{Lide_CRC_2004}         &  45.6(45.5)\cite{IhsanBarin} \\
  \ce{As1B1}       & 181292   &  126.6                                                 & 0.80                                           &    13.0                                   &  34.9                                &
  \ce{In1Te1}      &  169431  &   34.0                                                 & 2.57                                           &    96.1                                   &  52.9(47.7)\cite{IhsanBarin} \\
  \ce{As1Ba1Li1}   & 56445    &   31.1                                                 & 1.27                                           &    55.7                                   &  70.9                                &
  \ce{In1Te1}      &  640622  &   32.3                                                 & 2.57                                           &    101.1                                  &  53.1(47.7)\cite{IhsanBarin} \\
  \ce{As1Ga1}      & 53964    &   57.2(74.8)\cite{Lam_prb_1987}                        & 0.76(0.75)\cite{slack}                         &    21.6(16.2)\cite{Lide_CRC_2004}         &  46.9(46.8)\cite{IhsanBarin}         &
  \ce{K1}          &  641218  &    2.6(3.1)                                            & 0.80                                           &   158.2                                   &  25.7(29.4)\cite{IhsanBarin} \\
  \ce{As1In1}      & 165462   &   43.6(58.0)\cite{Lam_prb_1987}                        & 0.64(0.57)\cite{slack}                         &    19.7(14.1)\cite{Lide_CRC_2004}         &  43.6(47.8)\cite{IhsanBarin}         &
  \ce{K2O1}        &  44674   &   45.1                                                 & 1.76                                           &   73.7                                    &  73.7 \\
  \ce{B1Sb1}       & 184571   &   94.9                                                 & 0.85                                           &    15.2                                   &  38.0                                &
  \ce{K2S1}        &  183837  &   35.9                                                 & 1.54                                           &    50.0                                   &  74.3 \\
  \ce{B2Ti1}       & 78847    &  231.4                                                 & 1.27                                           &    15.3                                   &  45.7                                &
  \ce{Li1Pd1}      &  642257  &   52.5                                                 & 1.62                                           &    80.9                                   &  45.5 \\
  \ce{B2V1}        & 167794   &  265.9                                                 & 1.35                                           &    15.8                                   &  45.8                                &
  \ce{Li1Pt1}      &  104777  &  110.2                                                 & 1.44                                           &    34.1                                   &  44.5 \\
  \ce{Be1}         & 52708    &  99.4                                                  & 1.63                                           &    54.2                                   &  16.5(16.52)\cite{IhsanBarin}        &
  \ce{Li2O1}       &  60431   &   72.8(88.0)                                           & 1.19                                           &    55.1                                   &  53.9 \\
  \ce{Be1Rh1}      & 58734    &  210.0                                                 & 2.10                                           &    32.5                                   &  43.1                                &
  \ce{Li2S1}       &  657596  &   31.3                                                 & 1.22                                           &   83.3                                    &  64.4 \\
  \ce{Be1S1}       & 186889   &   89.7(105.0)\cite{Xu_sr_2013}                         & 0.91                                           &    20.9                                   &  36.7(34.2)\cite{IhsanBarin}         &
  \ce{Li2Se1}      &  168446  &   28.4                                                 & 1.04                                           &    70.7                                   &  66.8 \\
  \ce{Be1Se1}      & 616419   &   70.8(92.0)\cite{Xu_sr_2013}                          & 0.80                                           &    21.1                                   &  39.9                                &
  \ce{Li2Te1}      &  642399  &   24.0                                                 & 1.06                                           &    69.4                                   &  69.0 \\
  \ce{Be1Te1}      & 290008   &   53.8(67.0)\cite{Xu_sr_2013}                          & 0.72                                           &    19.9                                   &  41.6                                &
  \ce{Mg1O1}       &  159372  &  147.3(164)\cite{Sumino_jpearth_1976}                  & 1.53(1.44)\cite{slack}                         &    33.3                                   &  38.3(37.2)\cite{IhsanBarin} \\
  \ce{Be2C1}       & 616184   &  179.3                                                 & 1.17                                           &    20.2                                   &  38.9(43.3)\cite{IhsanBarin}         &
  \ce{Mg1Pt3}      &  104857  &  164.1                                                 & 2.70                                           &    40.1                                   &  98.4 \\
  \ce{Bi1Na1}      & 616837   &   24.1                                                 & 1.92                                           &    106.8                                  &  51.9                                &
  \ce{Mg1S1}       &  53939   &   68.1                                                 & 1.68                                           &    50.7                                   &  46.4(45.6)\cite{IhsanBarin} \\
  \ce{Br1Cu1}      & 30090    &   36.8                                                 & 1.19                                           &    53.5(46.2)\cite{Lide_CRC_2004}         &  49.5(54.8)\cite{IhsanBarin}         &
  \ce{Mg1Sc1}      &  108583  &   48.1                                                 & 1.56                                           &    51.3                                   &  47.8 \\
  \ce{Br1K1}       & 52243    &   11.0(15.4)\cite{Hauss_zfp_1960}                      & 1.87(1.45)\cite{slack}                         &   170.5(116.1)\cite{Li_jpcrd_1976}        &  53.7(52.3)\cite{Li_jpcrd_1976}      &
  \ce{Mg1Se1}      &  159398  &   53.3                                                 & 0.27                                           &    71.8                                   &  46.2 (48.0)\\
  \ce{Br1Li1}      & 53819    &   19.4(26.0)\cite{Hauss_zfp_1960}                      & 2.39                                           &   216.6(149.4)\cite{Li_jpcrd_1976}        &  54.9(48.9)\cite{Li_jpcrd_1976}      &
  \ce{Mg2Pb1}      &  104846  &   29.9                                                 & 1.27                                           &    59.4                                   &  73.1(72.5)\cite{IhsanBarin} \\
  \ce{Br1Na1}      & 44278    &   16.0(19.6)\cite{Hauss_zfp_1960}                      & 1.69(1.5)\cite{slack}                          &   148.0(126.9)\cite{Li_jpcrd_1976}        &  52.3(51.9)\cite{Li_jpcrd_1976}      &
  \ce{Mg2Si1}      &  163708  &   17.0                                                 & 1.22                                           &    44.3(34.5)\cite{Lide_CRC_2004}         &  68.7(68.0)\cite{IhsanBarin} \\
  \ce{C1}          & 182729   &  416.9(442.0)\cite{Madelung_Semiconductors_2004}       & 0.73(0.75)\cite{slack}                         &     3.2(3.5)\cite{Lide_CRC_2004}         &   6.4 (8.6)\cite{IhsanBarin}          &
  \ce{Mg2Sn1}      &  151368  &   37.8                                                 & 1.30                                           &    50.0(29.7)\cite{Lide_CRC_2004}         &  72.3 \\
  \ce{C1Si1}       & 618777   &  211.0(211.0)\cite{Madelung_Semiconductors_2004}       & 0.72(0.75)\cite{slack}                         &     7.3                                   &  27.7(27.1)\cite{IhsanBarin}         &
  \ce{Mn1O1}       &  18006   &  136.3                                                 & 1.78                                           &    39.9                                   &  44.1(44.17)\cite{IhsanBarin} \\
  \ce{C1Ti1}       & 181681   &  230.0                                                 & 1.44                                           &    16.8                                   &  35.1(33.9)\cite{IhsanBarin}         &
  \ce{Mn1S1}       &  76205   &   52.9                                                 & 0.42                                           &    12.8                                   &  45.9(45.6)\cite{IhsanBarin} \\
  \ce{C1Zr1}       & 180599   &  212.0                                                 & 1.58                                           &    17.9                                   &  38.7                                &
  \ce{Mn1Se1}      &  24252   &   45.0                                                 & 0.30                                           &    9.7                                    &  47.6(51.0)\cite{IhsanBarin} \\
  \ce{Ca1Cd1}      & 619188   &   27.3                                                 & 1.57                                           &    78.6                                   &  50.4                                &
  \ce{Mn1Te1}      &  181324  &   68.9                                                 & 0.23                                           &    7.9                                    &  48.3 \\
  \ce{Ca1F2}       & 40938    &   68.3(84.0)                                           & 1.78                                           &    68.7                                   &  68.7(68.7)\cite{IhsanBarin}         &
  \ce{N1Sc1}       &  155049  &  179.2                                                 & 1.57                                           &    22.7                                   &  38.6 \\
  \ce{Ca1O1}       & 180198   & 98.6(113)\cite{Chang_jphyschemsolids_1977}             & 1.57(1.57)\cite{slack}                         &    39.3                                   &  43.4(42.2)\cite{IhsanBarin}         &
  \ce{N1Ti1}       &  183415  &  248.3                                                 & 1.71                                           &    21.0                                   &  37.6(37.2)\cite{IhsanBarin} \\
  \ce{Ca1S1}       & 619530   &   69.6                                                 & 1.46                                           &    34.1                                   &  47.0(47.4)\cite{IhsanBarin}         &
  \ce{Na1}         &  644903  &    6.7                                                 & 1.32                                           &   20.3                                    &  26.4 \\
  \ce{Ca1Se1}      & 619570   &   46.0                                                 & 1.50                                           &    47.8                                   &  48.8(48.1)\cite{IhsanBarin}         &
  \ce{Se1Zn1}      &  181761  &  53.3                                                  & 0.83                                           &    26.4                                   &  47.8\\
  \ce{Ca1Te1}      & 619616   &   34.6                                                 & 1.47                                           &    51.3                                   &  49.5(41.6)\cite{IhsanBarin}         &
  \ce{Ni1O1}       &  166115  &  168.9                                                 & 1.59                                           &    33.6                                   &  42.1(44.5)\cite{IhsanBarin} \\
  \ce{Cd1F2}       & 28864    &   79.3                                                 & 2.20                                           &    74.1                                   &  71.2                                &
  \ce{Ni1Sb1}      &  646431  &  107.3                                                 & 1.73                                           &    36.2                                   &  48.6 \\
  \ce{Cd1O1}       & 181735   &  114.4                                                 & 1.94                                           &    46.0                                   &  46.1(43.7)\cite{IhsanBarin}         &
  \ce{Ni1Sc1}      &  105333  &  96.2                                                  & 1.86                                           &    45.6                                   &  48.7 \\
  \ce{Cd1Pd1}      & 620270   &  91.6                                                  & 2.32                                           &    57.1                                   &  50.4                                &
  \ce{Ni1Zn1}      &  647134  &  129.5                                                 & 1.86                                           &    46.8                                   &  48.1 \\
  \ce{Cd1Pt1}      & 620297   &  119.3                                                 & 2.70                                           &    51.7                                   &  50.7                                &
  \ce{Ni2Sn1Ti1}   &  646777  &  138.5                                                 & 2.17                                           &    41.3                                   &  97.2(98.0)\cite{IhsanBarin} \\
  \ce{Cd1S1}       & 290009   &   46.8                                                 & 0.53(0.75)\cite{Morelli_Slack_2006}            &    16.8                                   &  46.7                                &
  \ce{O1Pd1}       &  26598   &  142.3                                                 & 1.51                                           &    19.8                                   &  39.1 \\
  \ce{Cd1Sr1}      & 102066   &   19.8                                                 & 1.51                                           &    89.0                                   &  51.3                                &
  \ce{O1Sr1}       &  26960   &   76.4(91.2)\cite{Chang_jphyschemsolids_1977}          & 1.63(1.52)\cite{slack}                         &    44.8                                   &  46.2(45.4)\cite{IhsanBarin} \\
  \ce{Cd3Zr1}      & 102093   &   71.8                                                 & 1.99                                           &    50.5                                   &  100.2                               &
  \ce{O1Zn1}       &  182356  &  121.0                                                 & 0.66(0.75)\cite{Morelli_Slack_2006}            &    15.7                                   &  40.3(41.17)\cite{IhsanBarin} \\
  \ce{Cl1Cu1}      & 78270    &   35.6                                                 & 0.86                                           &    46.2(36.3)\cite{Lide_CRC_2004}         &  48.3(52.6)\cite{IhsanBarin}         &
  \ce{O1Zn1}       &  647683  &  121.4(143)\cite{Madelung_Semiconductors_2004}         & 0.66                                           &    15.7                                   &  40.3 \\
  \ce{Cl1K1}       & 240522   &   13.3(18.2)\cite{Hauss_zfp_1960}                      & 1.77(1.45)\cite{slack}                         &   157.4(111.3)\cite{Li_jpcrd_1976}        &  52.7(51.7)\cite{Li_jpcrd_1976}      &
  \ce{Pd1Zn1}      &  649134  &  115.0                                                 & 2.43                                           &    58.4                                   &  50.0 \\
  \ce{Cl1Li1}      & 26909    &   38.9(32.9)\cite{Hauss_zfp_1960}                      & 2.71                                           &   119.6(131.4)\cite{Li_jpcrd_1976}        &  50.0(48.1)\cite{Li_jpcrd_1976}      &
  \ce{Pt1Zn1}      &  105852  &  159.4                                                 & 2.13                                           &    37.6                                   &  49.2 \\
  \ce{Cl1Na1}      & 240600   &   19.7(25.1)\cite{Hauss_zfp_1960}                      & 1.75(1.56)\cite{slack}                         &   145.7(119.1)\cite{Li_jpcrd_1976}        &  51.6(50.54)\cite{Li_jpcrd_1976}     &
  \ce{S1Zn1}       &  108733  &   65.6(71.4)\cite{Lam_prb_1987}                        & 0.79(0.75)\cite{slack}                         &    23.1                                   &   45.5 \\
  \ce{Cl1Rb1}      & 18016    &   10.2(16.2)\cite{Hauss_zfp_1960}                      & 1.55(1.45)\cite{slack}                         &   156.5(108.0)\cite{Li_jpcrd_1976}        &  52.6(52.3)\cite{Li_jpcrd_1976}      &
  \ce{Sc1}         &  164093  &   52.1                                                 & 1.10                                           &    30.8                                   &  23.8(25.58)\cite{IhsanBarin} \\
  \ce{Cu1I1}       & 163427   &   18.9                                                 & 1.36                                           &   101.0                                   &  51.0                                &
  \ce{Sc1Zn1}      &  106041  &   65.1                                                 & 1.39                                           &    42.3                                   &  48.3 \\
  \ce{Cu1}         & 627117   &  114.5                                                 & 1.91                                           &    53.5                                   &  24.3(24.5)\cite{IhsanBarin}         &
  \ce{Si1}         &  76268   &   84.9(100.0)\cite{Madelung_Semiconductors_2004}       & 0.42(0.56)\cite{Morelli_Slack_2006}            &     8.0(7.8)\cite{Lide_CRC_2004}          &  20.0(20.0)\cite{IhsanBarin} \\
  \ce{Cu1Sn1}      & 629278   &   54.7                                                 & 2.40                                           &    85.4                                   &  51.9                                &
  \ce{Tc1V1}       &  106143  &  232.5                                                 & 1.91                                           &    21.7                                   &  45.9 \\
  \ce{Cu2Ni1Zn1}   & 103079   &  124.7                                                 & 1.93                                           &    50.5                                   &  96.7                                &
  \ce{Te1Zn1}      &  184485  &   39.4(51.0)\cite{Lam_prb_1987}                        & 0.89(0.97)\cite{slack}                         &    30.7                                   &  48.6 \\
  \ce{F1K1}        & 52241    &   22.5(31.6)\cite{Hauss_zfp_1960}                      & 1.59(1.5)\cite{slack}                          &   133.4(104.4)\cite{Li_jpcrd_1976}        &  50.5(49.0)\cite{Li_jpcrd_1976}      &
  \ce{Te1Zr1}      &  653209  &   93.1                                                 & 1.60                                           &    27.6                                   &  48.2 \\
  \ce{F1Li1}       & 53839    &   55.8(76.0)\cite{Hauss_zfp_1960}                      & 1.63(1.5)\cite{slack}                          &    109.9(99.6)\cite{Li_jpcrd_1976}        &  42.8(42.0)\cite{Li_jpcrd_1976}      &
  \ce{Ti1}         &  168830  &  109.3                                                 & 1.62                                           &    30.1                                   &  23.5(25.1)\cite{IhsanBarin} \\
  \ce{F1Na1}       & 52238    &   37.3(48.5)\cite{Hauss_zfp_1960}                      & 1.52(1.5)\cite{slack}                          &    113.7(95.1)\cite{Li_jpcrd_1976}        &  47.8(46.9)\cite{Li_jpcrd_1976}      &
  \ce{Zn1}         &  181734  &   44.8                                                 & 2.47                                           &   144.8                                   &  26.8(25.41)\cite{IhsanBarin} \\
  \ce{Ga1P1}       & 77088    &   74.8(88.7)\cite{Lam_prb_1987}                        & 0.70(0.75)\cite{slack}                         &    16.2(15.9)\cite{Lide_CRC_2004}         &  44.1(44.2)\cite{IhsanBarin}         &
  \ce{Zn1Zr1}      &  181290  &   96.3                                                 & 1.42                                           &    31.2                                   &  48.3 \\
  \ce{Ga1Sb1}      & 41675    &   42.7(45.1)\cite{Lam_prb_1987}                        & 0.71(0.75)\cite{slack}                         &    21.9                                   &  47.9                                &
                   &          &                                                        &                                                &                                           &                                      \\
\label{tab:data}
\end{longtable}
\end{scriptsize}
\end{landscape}

\begin{figure}[h!]
  \centerline{\includegraphics*[width=0.50\textwidth,clip=true]{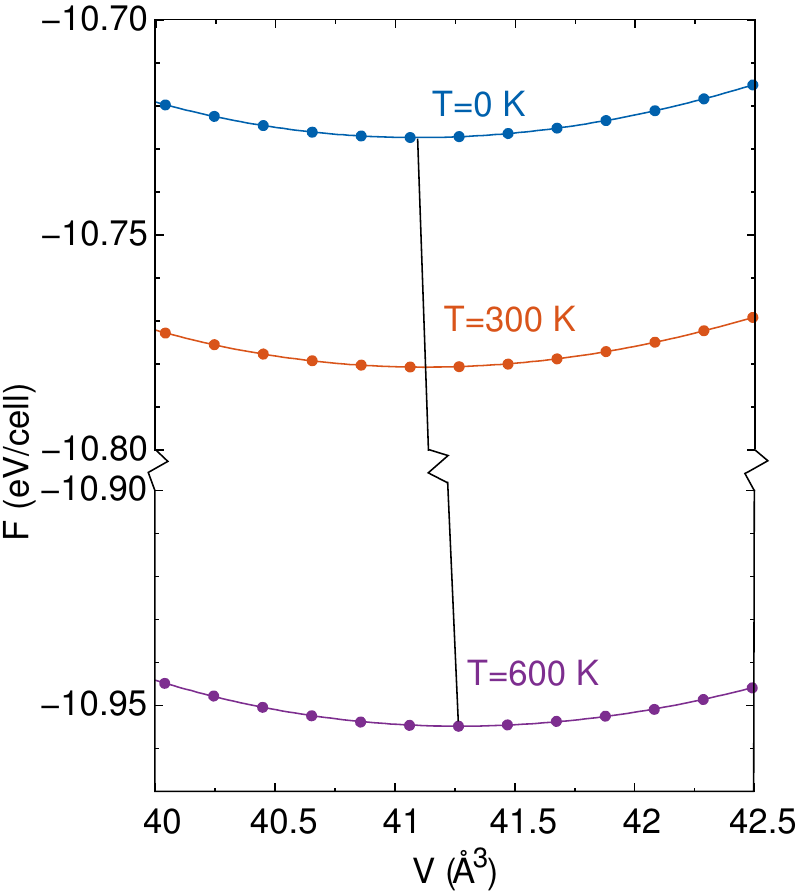}}
\vspace{-1mm}
  \caption{\small Free energy and volume data fitted using Birch-Murnaghan function for different temperatures for Si. Points represent calculated data and solid lines indicate the fitted function. Equilibrium volumes at each temperature are connected by solid black line.}
  \label{fig:VolumeFree}
\end{figure}

\begin{figure}[h!]
  \centerline{
    \includegraphics*[width=\textwidth,clip=true]{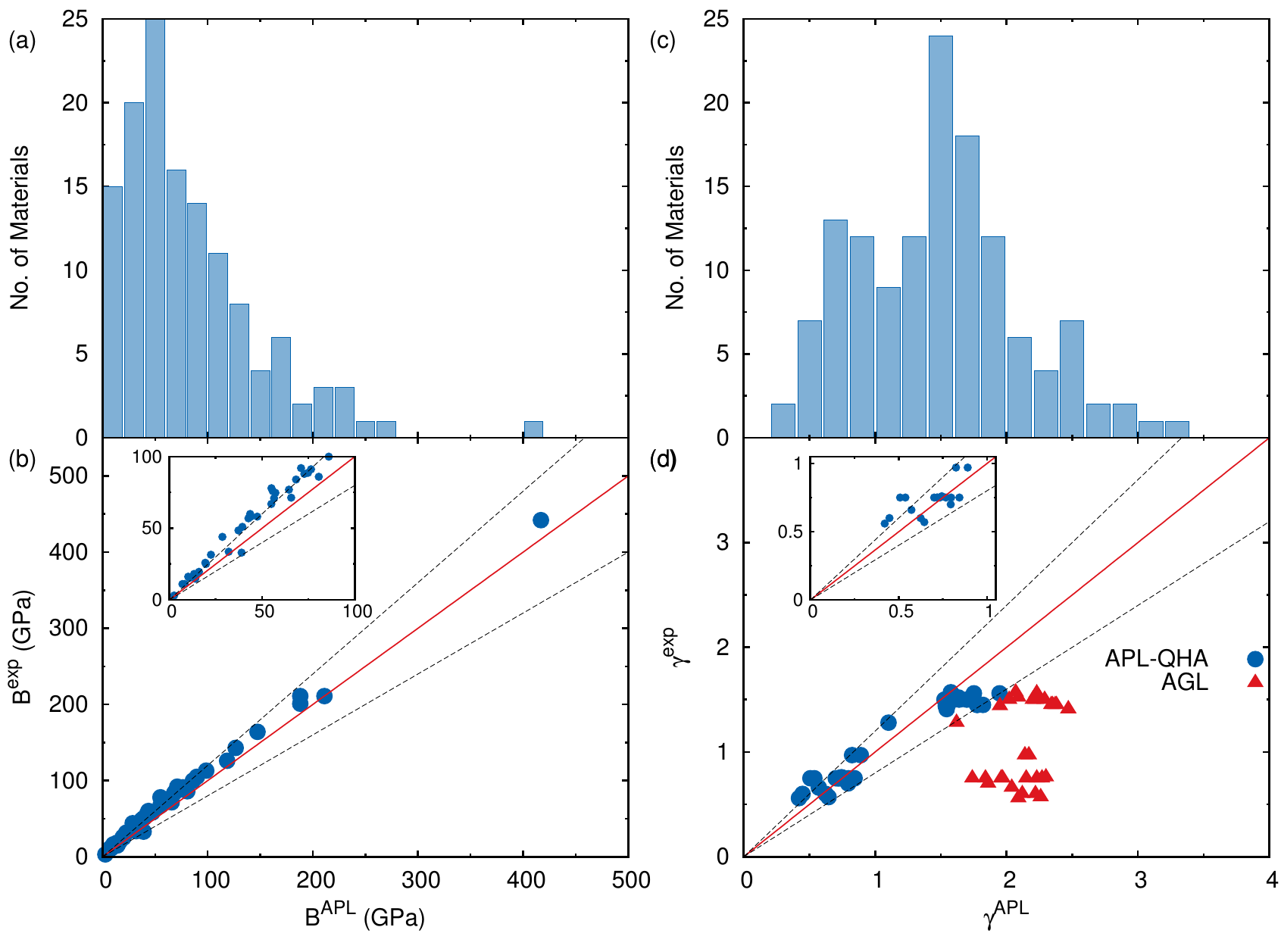}
  }
  \vspace{-1mm}
  \caption{\small
    {{
    {\bf (a)} Histogram for the dispersion of predicted $B$ at 300 K for the selected set of systems. 
    {\bf (b)} Comparison of experimental $B^{exp}$ and predicted $B^{APL}$ for the selected set of systems. The red line represents equality between experimental and predicted values  and the dashed black lines represent $\pm$20\% error.
    {\bf (c)} Histogram for the dispersion of predicted Gr{\"u}neisen parameters at 300 K for the selected set of systems. 
    {\bf (d)} Comparison of experimental $\gamma^{exp}$ and predicted  $\gamma^{APL}$ for the selected set of systems. 
    Ther red line represents equality between experimental and predicted values  and the dashed black lines represent $\pm$20\% error. The blue circles represent QHA-APL  results and the red triangles correspond to AGL results \cite{curtarolo:art96}.
  }}}
  \label{fig:BulkModulus}
  \label{fig:Gruneisen}
\end{figure}

For each temperature, we calculated the free energy ($F$)  at different volumes (V).
These $F$ vs. $V$ curves are initially fitted to a cubic function which is used as an initial guess to fit the Birch-Murnaghan equation of state (see ``Method'' and Eq. \ref{EQ:BM}). As an example, the results of the final fitting for Si are depicted in Fig.~\ref{fig:VolumeFree}.
Automatic tests to ensure the correct fitting of the curves have been implemented to warn of possible errors, especially in magnetic systems or close to phase transition temperatures.
The bulk modulus, $B$, at 300 K is obtained by the fitting procedure.
Fig. \ref{fig:BulkModulus} (a) illustrates the distribution of $B$ for the whole set of materials while Fig. \ref{fig:BulkModulus} (b) compares experimental and  predicted  values of $B$ for those materials with available experimental data.
 The computed values range over two orders of magnitude, from C diamond (442 GPa) to K (3.2 GPa). 
To quantify the accuracy, precision and robustness of our results we have used different statistical quantities such as mean absolute deviation (MAD), root mean square deviation (RMSD), root mean square relative deviation (RMSrD), relative maximum absolute deviation (rMAX), and Pearson and Spearman correlation (see Table \ref{tab:stats}). 
Most of the predicted values are within a 20\% error (dashed lines in Fig. \ref{fig:BulkModulus} (b)), presenting a RMSrD of around 17\% and a rMAX close to 22.9\%. 
Values close to 1.0 for the Pearson and Spearman correlations also demonstrate that our implementation of the QHA is robust and that the results can be compared with experimental data.

\begin{table}
  \caption{\small Mean absolute deviation (MAD), root mean square deviation (RMSD), root mean square relative deviation (RMSrD), relative maximum absolute deviation (rMAX), and Pearson and Spearman correlation for the material data set. Units: MAD and RMSD in the same units that are used in Table \ref{tab:data}}
  \centering
  \begin{tabular}[b]{c C{1.2cm} C{1.2cm} C{1.2cm} C{1.2cm} }
    \hline \hline
    & $B$ & $\gamma$ & $\alpha$ & $C_p$  \\
    \hline
    MAD          & 11.09  &  0.13   &  20.22  &  1.63      \\
    RMSD         & 12.89 &  0.17   & 28.15  &  2.45      \\
    RMSrD        & 17.38\% &  15.00\%   & 28.0\%  &  6.35\%      \\
    rMAX         & 22.9\%  &  28.32\%   & 46.5\%   & 18.9\%       \\
    $r$          & 0.996 &   0.969 &  0.974 &  0.985     \\
    $\rho$       & 0.990 &   0.891 &  0.927 &  0.929     \\
    \hline \hline
  \end{tabular}
  \label{tab:stats}
\end{table}

In the spirit of the QHA, the Gr{\"u}neisen parameter, $\gamma(T)$, is often used to estimate the anharmonicity of the vibrations in the crystal. 
Methods such as AGL are able to predict reasonably accurate values for the Debye temperature or $B$ but fail to predict $\gamma(T)$ \cite{curtarolo:art96}. 
Our QHA-APL results are shown in Fig. \ref{fig:Gruneisen} which includes, for comparison, the available results using AGL for the same compounds.  
Despite the higher computational cost of the QHA compared to the AGL methodology (one or two orders of magnitude higher than AGL depending on the size of the system), it is clear that results are tremendously improved. 
The MAD for QHA-APL is 0.13 for a property whose highest measured value is around 3 and the average values are between 1 and 2.
 Relative statistical indicators also demonstrate the high accuracy of the method obtaining a RMSrD below 16\% and a rMAX lower than 29\%. 
Values larger than 3 are especially interesting for their potential application as thermoelectrics.
 As we mentioned, $\gamma(T)$ is related to the anharmonicity of the crystal, so high values for this property usually \cite{curtarolo:art84} 
 indicate lower values of the lattice thermal conductivity, $\kappa_L$, which increases the thermoelectric figure of merit, $ZT$.
We have found two  materials with $\gamma$ over 3, however both  are metals (HgPd, HgPt). We have looked for other possible candidates with high values of $\gamma$ considering also that for a high thermoelectric figure of merit they should be semiconductors. K$_2$O and MnO are the best two candidates, with  $\gamma$ around 1.8 and band gaps below 2 eV.  
There is no experimental data available for K$_2$O, although Gheribi \textit{et al}. have predicted a lattice thermal conductivity for this material below 2 Wm$^{-1}$K$^{-1}$ \cite{Gheribi_jap_2015}.

\begin{figure*}[h!]
  \centerline{\includegraphics*[width=0.95\textwidth,clip=true]{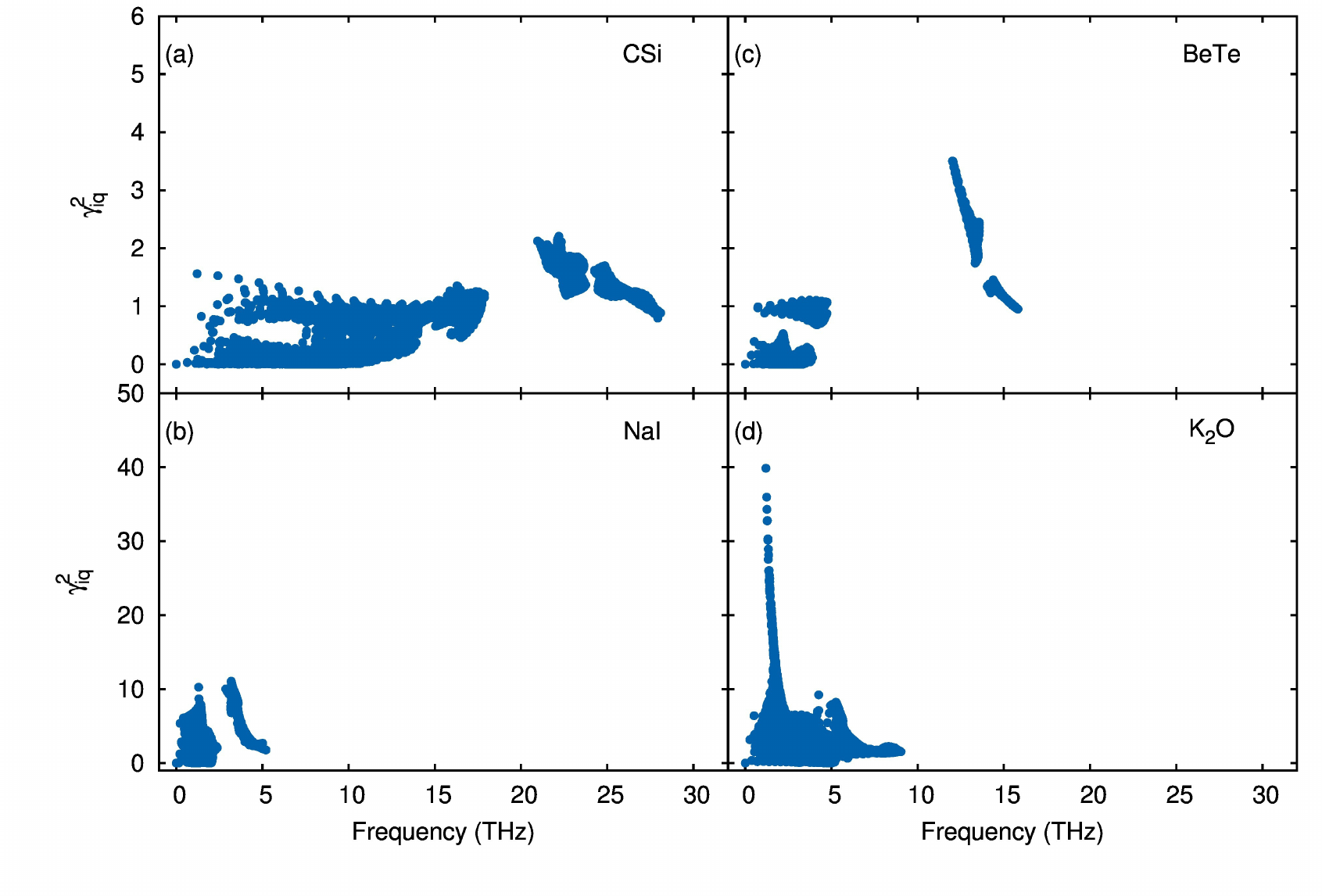}}
\vspace{-1mm}
  \caption{\small Mode Gr{\"u}neisen parameters for {\bf (a)} SiC, {\bf (b)} NaI, {\bf (c)} BeTe, and {\bf (d)} \ce{K2O}.}
  \label{fig:MGru}
\end{figure*}

The Gr{\"u}neisen parameters for each vibrational mode, $\gamma_{iq}$, provide even more information than $\gamma$ about the anharmonicity of the crystal and the lattice thermal conductivity.
High and low values of $\kappa_L$ in an insulator are usually linked to low and high values of $\gamma_{iq}$ at low frequency. 
SiC ($\kappa$=360 Wm$^{-1}$K$^{-1}$)~\cite{Morelli_Slack_2006} and NaI ($\kappa$=1.8 Wm$^{-1}$K$^{-1}$)~\cite{Morelli_Slack_2006} are two good examples of this trend (see Fig. \ref{fig:MGru} (a) and (b)) that has been exploited in the literature to pinpoint anharmonic effects \cite{Orabi_cm_2016,Vaqueiro_pccp_2015}.
We have studied the mode Gr{\"u}neisen parameters of two other materials included in the benchmark whose $\kappa_L$ has not been measured experimentally.
We have already discussed the probable low thermal conductivity of \ce{K2O} which is in agreement with the very high mode Gr{\"u}neisen parameter values at low frequencies (see Fig.~\ref{fig:MGru} (d)).
BeTe shows the opposite behavior to \ce{K2O}, presenting low values for $\gamma_{iq}$ at low frequencies (see Fig.~\ref{fig:MGru} (c)).
This trend is in agreement with the AGL prediction of $\kappa_L$ for BeTe, which is close to the value obtained for \ce{AlAs} ($\kappa$=98 Wm$^{-1}$K$^{-1}$)~\cite{Morelli_Slack_2006}.
These predictions validate the use of $\gamma$ and  $\gamma_{iq}$ as a simple predictors for $\kappa_L$ and demonstrates that the QHA-APL method  can be used as a powerful method for the discovery of new interesting properties in well known materials.

\begin{figure}[h!]
  \centerline{
    \includegraphics*[width=\textwidth,clip=true]{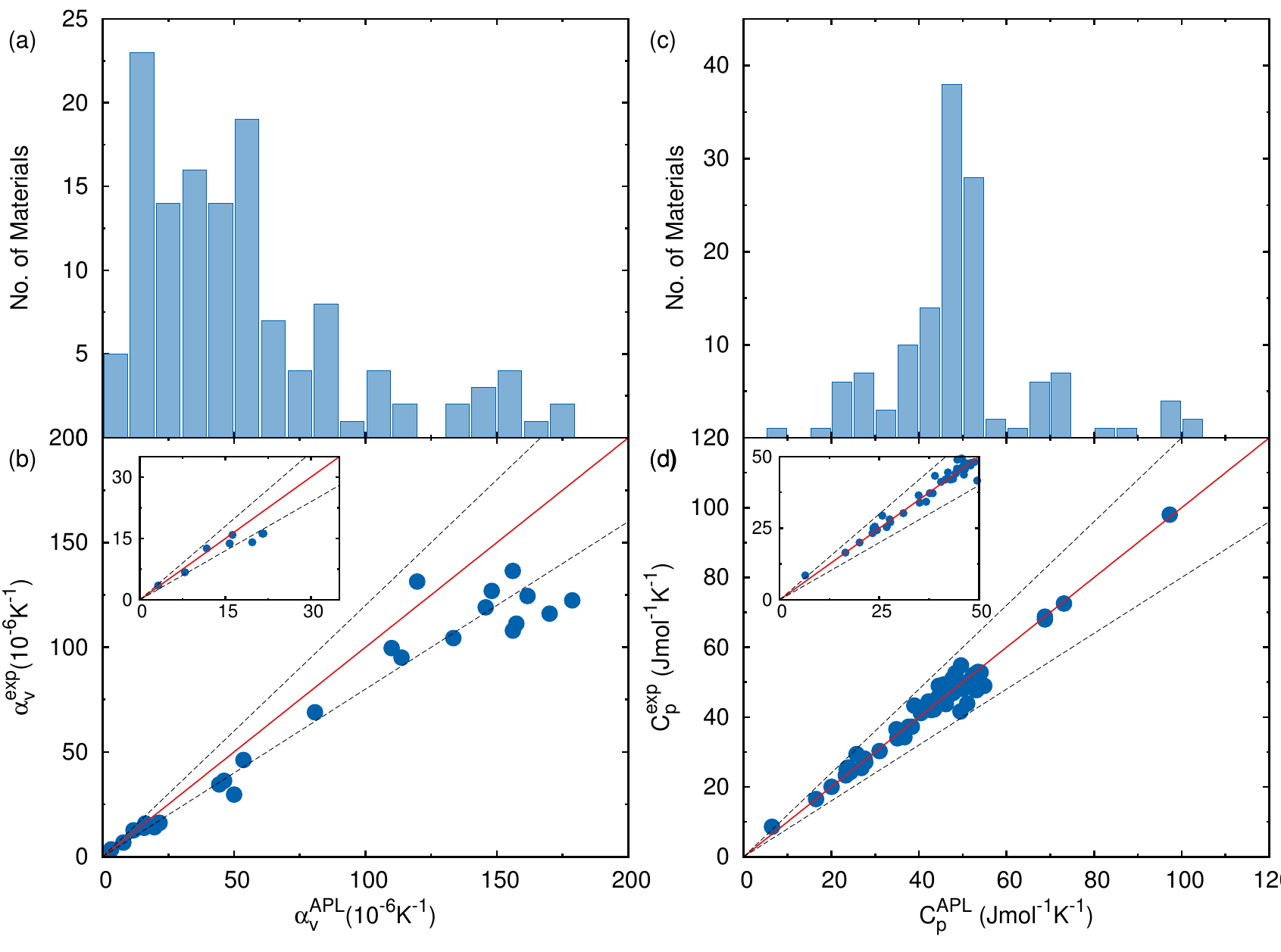}
  }
  \vspace{-1mm}
  \caption{\small 
    {{
    {\bf (a)} Histogram for the dispersion of predicted volumetric thermal expansion coefficient at 300 K for the selected set of systems. 
    {\bf (b)} Comparison of experimental $\alpha^{exp}$ and predicted $\alpha^{APL}$  for the selected set of systems. The red line represents equality between experimental and predicted values  and the dashed black lines represent $\pm$20\% error.
    {\bf (c)} Histogram for the dispersion of predicted isobaric specific heat at 300 K for the selected set of systems.
    {\bf (d)} Comparison of experimental $C_p^{exp}$ and predicted $C_p^{APL}$ for the selected set of systems. The red line represents equality between experimental and predicted values  and the dashed black lines represent $\pm$20\% error.
  }}}
  \label{fig:CTE}
  \label{fig:CP}
\end{figure}

The volumetric thermal expansion, $\alpha_V(T)$, has very important  implications in engineering. However, it is not easy to predict accurately from first principles. Statistical parameters prove again that QHA-APL methodology is a reliable method to obtain this thermodynamic quantity. The uncertainty as described by RMSrD is below 28\% and MAD is around 20.22 10$^{-6}$K$^{-1}$ for the benchmark in which the range varies between 0 and 160 10$^{-6}$K$^{-1}$. Low thermal expansion coefficient materials deserve special attention because they are particularly interesting for a variety of applications. There are 5 materials in our benchmark that present a value of $\alpha$ below 10$^{-5}$K$^{-1}$ (see Fig. \ref{fig:CTE} (b)). Three of them have been already reported, C, Si and CSi. However, to the best of our knowledge, the other two materials have not been experimentally measured or predicted: MnTe and MgSe.

Isobaric specific heat, $C_p$, can be considered another good test for our method because of the vast amount of available experimental data (see Fig. \ref{fig:CP} (d)).
The low MAD (9.84 Jmol$^{-1}$K$^{-1}$) is surprising  for a property in which the average value of our data set is close to 50 Jmol$^{-1}$K$^{-1}$. 
This fact is also reflected in the relative statistical parameters (RMSrD = 6.35\% and rMAX = 18.9\%) and correlations close to 1.0.
A caveat should be mentioned: this deviation grows slightly when  comparing $C_p$ at 1000 K instead of 300 K. This is due to higher anharmonicity at higher temperatures.

Some groups  of materials  are more likely to show larger deviations because of the  exchange-correlation functional limitations. Inaccurate band gaps could lead to predict metallic behavior in semiconductor with a narrow band gap. This qualitative disagreement can modify the electronic terms in  
the free energy. To ensure that our description 
is correct, we have calculated the RMSrD for metals and semiconductors with a predicted
band gap smaller than 1.0 eV. For metals, the RMSrD is 8.1\% for $C_p$, which is below the RMSrD for the whole data set. We have also calculated this quantity for $\alpha_V(T)$ and $C_p$ for low band gap semiconductors, obtaining 30.0\% and 3.6\% respectively. Both values are also close to the RMSrD obtained for the  whole data set.
Despite QHA limitations, we demonstrate that our implementation can predict the properties of different groups of materials.

\begin{figure}[h!]
  \centerline{
    \includegraphics*[width=\textwidth,clip=true]{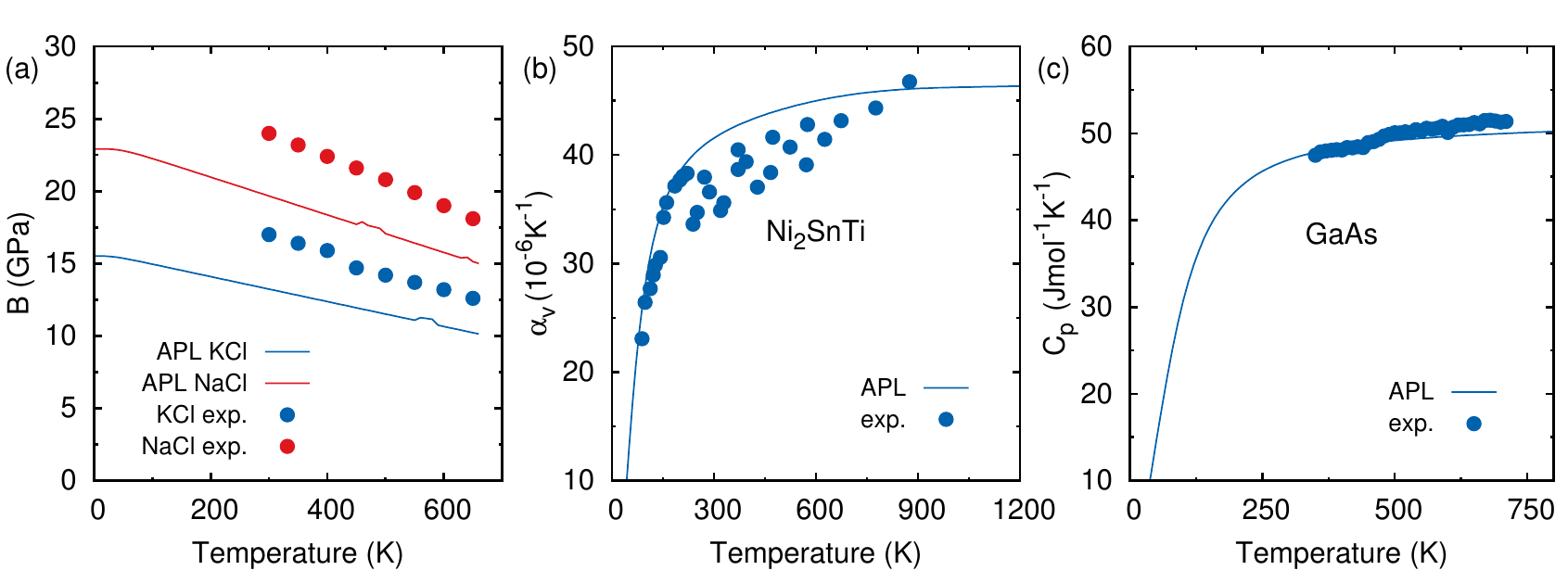}
  }
  \vspace{-1mm}
  \caption{\small 
    {\bf (a)} Predicted bulk modulus of \ce{NaCl} and \ce{KCl} compared to the available experimental data~\cite{SKSrivastavaIJoPaAP}. 
    {\bf (b)} Predicted volumetric thermal expansion coefficient of \ce{Ni2SnTi} compared to the available experimental data~\cite{PHermetJPhysChem}.
    {\bf (c)} Predicted isobaric specific heat of \ce{GaAs} compared to the available experimental data~\cite{Blakemore_jap1982}.
  }
  \label{fig:temp}
\end{figure}

We  also computed  the temperature dependence for selected
materials in the data set (see Fig.~\ref{fig:temp}). Our  implementation predicts correctly the behavior of $\gamma(T)$, $B$, $\alpha_V(T)$ and $C_p$  below melting points for various types of materials (metals, insulators and semiconductors). The bulk modulus for insulators such as \ce{NaCl} and \ce{KCl} are shown in Fig.~\ref{fig:temp} (a) with errors smaller than 25\%. QHA-APL  also predicts $\alpha_V(T)$ for complex metals as \ce{Ni2SnTi} with a high accuracy below its melting point (see Fig.~\ref{fig:temp} (b)). Finally, $C_p$ for \ce{GaAs}, a classic semiconductor, is depicted in Fig.~\ref{fig:temp} (c) presenting errors smaller than  2\%.

\begin{figure}[h!]
  \centerline{\includegraphics*[width=\textwidth,clip=true]{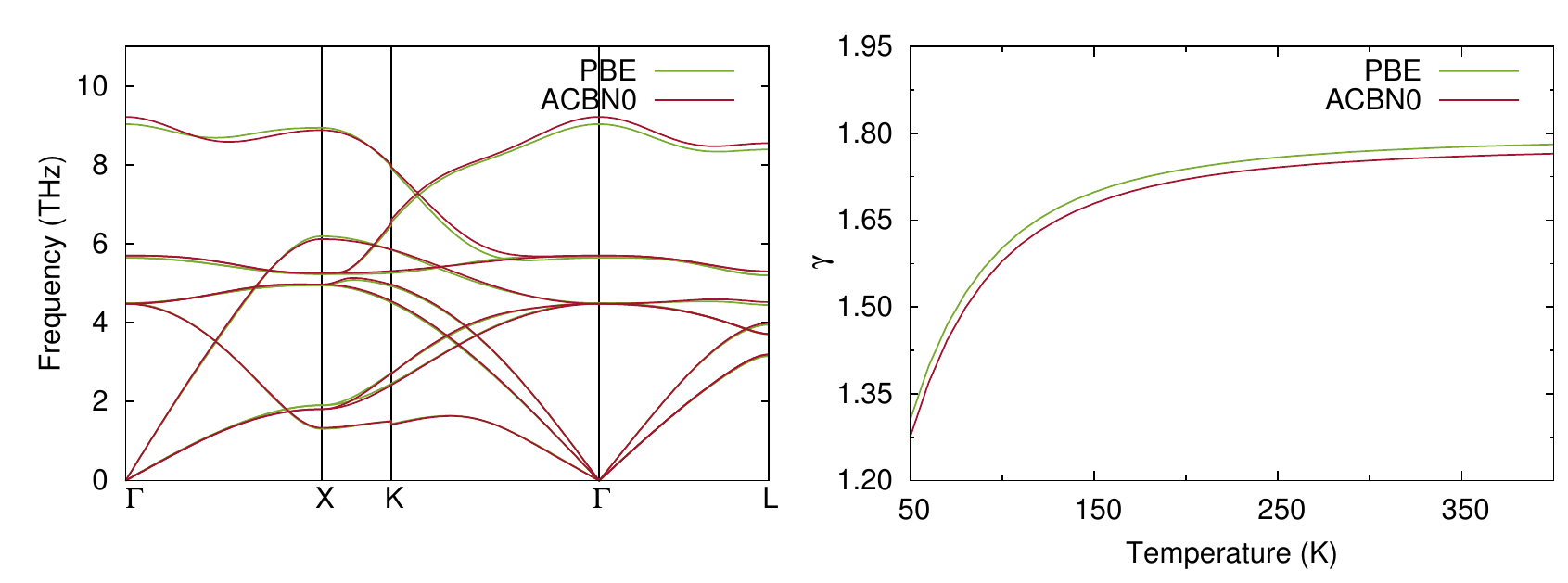}}
\vspace{-1mm}
  \caption{\small {\bf (a)} Phonon dispersion curves and {\bf (b)} average Gr{\"u}neisen parameter for \ce{K2O} using PBE and ACBN0 functionals.}
\label{fig:ACBN0}
\end{figure}

More sophisticated electronic structure methods can be used to validate our data. Our framework can use different exchange correlation functionals such as the pseudo hybrid functional ACBN0 \cite{curtarolo:art93}.
We have used \ce{K2O} as a test to study how a more accurate functional works compared to the results obtained with PBE (see Fig.~\ref{fig:ACBN0}). Despite the use of a more accurate approach, the band gap (1.8 eV compared to 1.71 eV for PBE), phonon dispersion curves and the average Gr{\"u}neisen parameter for \ce{K2O} are only slightly modified by ACBN0.

\section{Conclusions}\label{conclusions}

The quasi-harmonic approximation has been combined with the Automatic Phonon Library in the {\small AFLOW} high-throughput framework in order to develop a tool  to accurately and robustly predict some of the most important temperature dependent properties of solids such as $B$, $\gamma$, $\alpha$, and $C_p$. Lattice dynamics calculations at different volumes for fully relaxed structures are performed, managed, self-corrected, and post processed by the QHA-APL code, creating a unique workflow to  populate material databases such as {\small AFLOW}.
    A benchmark of 130 materials has been used to test the precision and accuracy of the software. QHA-APL predicts results whose RMSrD is below 18\% for any of the four quantities with respect to experimentally reported data. We confirm recent results on \ce{K2O} and computed data for well known materials that have not been reported before.
     QHA-APL is useful not just for the prediction of properties for particular materials; its automation, robustness, accuracy and precision make it the perfect framework for the creation of material data sets that can be used for data mining or machine learning.

\section{Acknowledgements}

We thank Dr. E. Perim, Dr. O. Levy, A. Supka, and M. Damian for various technical discussions. 
We would like to acknowledge support by DOD-ONR (N00014-13-1-0635, N00014-11-1-0136, N00014-09-1-0921)
and by DOE (DE-AC02- 05CH11231), specifically the BES program under Grant \# EDCBEE. 
The {\small AFLOW} consortium would like to acknowledge the Duke University - Center for Materials Genomics and the CRAY corporation for computational support.

\newpage

\bibliographystyle{elsarticle-num-names}

\begin{thebibliography}{65}
\providecommand{\natexlab}[1]{#1}
\providecommand{\url}[1]{\texttt{#1}}
\providecommand{\urlprefix}{URL }
\expandafter\ifx\csname urlstyle\endcsname\relax
  \providecommand{\doi}[1]{doi:\discretionary{}{}{}#1}\else
  \providecommand{\doi}[1]{doi:\discretionary{}{}{}\begingroup
  \urlstyle{rm}\url{#1}\endgroup}\fi
\providecommand{\bibinfo}[2]{#2}

\bibitem[{Curtarolo et~al.(2013)Curtarolo, Hart, {Buongiorno~Nardelli}, Mingo,
  Sanvito, and Levy}]{nmatHT}
\bibinfo{author}{S.~Curtarolo}, \bibinfo{author}{G.~L.~W. Hart},
  \bibinfo{author}{M.~{Buongiorno~Nardelli}}, \bibinfo{author}{N.~Mingo},
  \bibinfo{author}{S.~Sanvito}, \bibinfo{author}{O.~Levy}, \bibinfo{title}{The
  high-throughput highway to computational materials design},
  \bibinfo{journal}{Nat.\ Mater.} \bibinfo{volume}{12} (\bibinfo{year}{2013})
  \bibinfo{pages}{191--201}, \doi{\bibinfo{doi}{10.1038/nmat3568}}.

\bibitem[{Blanco et~al.(2004)Blanco, Francisco, and
  Lua{\~n}a}]{BlancoGIBBS2004}
\bibinfo{author}{M.~A. Blanco}, \bibinfo{author}{E.~Francisco},
  \bibinfo{author}{V.~Lua{\~n}a}, \bibinfo{title}{GIBBS: isothermal-isobaric
  thermodynamics of solids from energy curves using a quasi-harmonic Debye
  model}, \bibinfo{journal}{Computer Physics Communications}
  \bibinfo{volume}{158}~(\bibinfo{number}{1}) (\bibinfo{year}{2004})
  \bibinfo{pages}{57--72}, \doi{\bibinfo{doi}{10.1016/j.comphy.2003.12.001}}.

\bibitem[{Toher et~al.(2014)Toher, Plata, Levy, {de~Jong}, Asta,
  {Buongiorno~Nardelli}, and Curtarolo}]{curtarolo:art96}
\bibinfo{author}{C.~Toher}, \bibinfo{author}{J.~J. Plata},
  \bibinfo{author}{O.~Levy}, \bibinfo{author}{M.~{de~Jong}},
  \bibinfo{author}{M.~D. Asta}, \bibinfo{author}{M.~{Buongiorno~Nardelli}},
  \bibinfo{author}{S.~Curtarolo}, \bibinfo{title}{High-Throughput Computational
  Screening of thermal conductivity, {Debye} temperature and {Gr{\"u}neisen}
  parameter using a quasi-harmonic {Debye} Model}, \bibinfo{journal}{Phys.\
  Rev.\ B} \bibinfo{volume}{90} (\bibinfo{year}{2014}) \bibinfo{pages}{174107},
  \doi{\bibinfo{doi}{10.1103/PhysRevB.90.174107}}.

\bibitem[{Li et~al.(2014)Li, Carrete, Katcho, and Mingo}]{ShengBTE_2014}
\bibinfo{author}{W.~Li}, \bibinfo{author}{J.~Carrete}, \bibinfo{author}{N.~A.
  Katcho}, \bibinfo{author}{N.~Mingo}, \bibinfo{title}{{ShengBTE:} a solver of
  the {B}oltzmann transport equation for phonons}, \bibinfo{journal}{Comput.\
  Phys.\ Commun.} \bibinfo{volume}{185} (\bibinfo{year}{2014})
  \bibinfo{pages}{1747--1758}, \doi{\bibinfo{doi}{10.1016/j.cpc.2014.02.015}}.

\bibitem[{Togo et~al.(2015)Togo, Chaput, and Tanaka}]{Togo_prb_2015}
\bibinfo{author}{A.~Togo}, \bibinfo{author}{L.~Chaput},
  \bibinfo{author}{I.~Tanaka}, \bibinfo{title}{Distributions of phonon
  lifetimes in Brillouin zones}, \bibinfo{journal}{Phys.\ Rev.\ B}
  \bibinfo{volume}{91} (\bibinfo{year}{2015}) \bibinfo{pages}{094306},
  \doi{\bibinfo{doi}{10.1103/PhysRevB.91.094306}}.

\bibitem[{Plata et~al.(2016)Plata, Nath, Usanmaz, Carrete, Toher, Fornari,
  Nardelli, and Curtarolo}]{APL2_2016}
\bibinfo{author}{J.~J. Plata}, \bibinfo{author}{P.~Nath},
  \bibinfo{author}{D.~Usanmaz}, \bibinfo{author}{J.~Carrete},
  \bibinfo{author}{T.~Toher}, \bibinfo{author}{M.~Fornari},
  \bibinfo{author}{M.~B. Nardelli}, \bibinfo{author}{S.~Curtarolo},
  \bibinfo{title}{APL 2.0: An automatic high throughput phonon calculator for
  finite temperature properties}, \bibinfo{journal}{In preparation} .

\bibitem[{Ho and Taylor(1998)}]{Taylor1998}
\bibinfo{editor}{C.~Y. Ho}, \bibinfo{editor}{R.~E. Taylor} (Eds.),
  \bibinfo{title}{Thermal Expansion of Solids}, \bibinfo{publisher}{Asm.
  Intl.}, \bibinfo{year}{1998}.

\bibitem[{Baroni et~al.(2010)Baroni, Giannozzi, and Isaev}]{Baroni_rmg_2010}
\bibinfo{author}{S.~Baroni}, \bibinfo{author}{P.~Giannozzi},
  \bibinfo{author}{E.~Isaev}, \bibinfo{title}{Density-Functional Perturbation
  Theory for Quasi-Harmonic Calculations}, \bibinfo{journal}{Rev.\ Mineral\
  Geochem.} \bibinfo{volume}{71} (\bibinfo{year}{2010})
  \bibinfo{pages}{39--57}, \doi{\bibinfo{doi}{10.2138/rmg.2010.71.3}}.

\bibitem[{Carrier et~al.(2007)Carrier, Wentzcovitch, and
  Tsuchiya}]{Carrier_prb_2007}
\bibinfo{author}{P.~Carrier}, \bibinfo{author}{R.~Wentzcovitch},
  \bibinfo{author}{J.~Tsuchiya}, \bibinfo{title}{First-principles prediction of
  crystal structures at high temperatures using the quasiharmonic
  approximation}, \bibinfo{journal}{Phys.\ Rev.\ B} \bibinfo{volume}{76}
  (\bibinfo{year}{2007}) \bibinfo{pages}{064116},
  \doi{\bibinfo{doi}{10.1103/PhysRevB.76.064116}}.

\bibitem[{Alf\'e(2009)}]{Alfe_cpc_2009}
\bibinfo{author}{D.~Alf\'e}, \bibinfo{title}{PHON: A program to calculate
  phonons using the small displacement method}, \bibinfo{journal}{Comput.\
  Phys.\ Commun.} \bibinfo{volume}{180} (\bibinfo{year}{2009})
  \bibinfo{pages}{2622--2633}, \doi{\bibinfo{doi}{10.1016/j.cpc.2009.03.010}}.

\bibitem[{Duong et~al.(2011)Duong, Gibbons, Kinra, and
  Arroyave}]{Duong_jap_2011}
\bibinfo{author}{T.~Duong}, \bibinfo{author}{S.~Gibbons},
  \bibinfo{author}{R.~Kinra}, \bibinfo{author}{R.~Arroyave},
  \bibinfo{title}{Ab-initio aprroach to the electronic, structural, elastic,
  and finite-temperature thermodynamic properties of {Ti$_2$AX} ({A} = {Al} or
  {Ga} and {X} = {C} or {N})}, \bibinfo{journal}{J.\ Appl.\ Phys.}
  \bibinfo{volume}{110} (\bibinfo{year}{2011}) \bibinfo{pages}{093504},
  \doi{\bibinfo{doi}{http://dx.doi.org/10.1063/1.3652768}}.

\bibitem[{Wang et~al.(2014)Wang, Wang, Li, Li, and Zhou}]{Wang_jacers_2014}
\bibinfo{author}{J.~Wang}, \bibinfo{author}{J.~Wang}, \bibinfo{author}{A.~Li},
  \bibinfo{author}{J.~Li}, \bibinfo{author}{Y.~Zhou},
  \bibinfo{title}{Theoretical Study on the Mechanism of Anisotropic Thermal
  Properties of {Ti$_2$AlC} and {Cr$_2$AlC}}, \bibinfo{journal}{J.\ Am.\
  Ceramic.\ Soc.} \bibinfo{volume}{97} (\bibinfo{year}{2014})
  \bibinfo{pages}{1202--1208}, \doi{\bibinfo{doi}{10.1111/jace.12814}}.

\bibitem[{Huang et~al.(2016)Huang, Lu, Tennessen, and
  Rondinelli}]{Huang_cms_2016}
\bibinfo{author}{L.~F. Huang}, \bibinfo{author}{X.~Z. Lu},
  \bibinfo{author}{E.~Tennessen}, \bibinfo{author}{J.~M. Rondinelli},
  \bibinfo{title}{An efficient ab-initio quasiharmonic approach for the
  thermodynamics of solids}, \bibinfo{journal}{Comput.\ Phys.\ Commun.}
  \bibinfo{volume}{120} (\bibinfo{year}{2016}) \bibinfo{pages}{84--96},
  \doi{\bibinfo{doi}{10.1016/j.commatsci.2016.04.012}}.

\bibitem[{Togo and Tanaka(2015)}]{Togo_scrmat_2015}
\bibinfo{author}{A.~Togo}, \bibinfo{author}{I.~Tanaka}, \bibinfo{title}{First
  principles phonon calculations in materials science}, \bibinfo{journal}{Scr.\
  Mater.} \bibinfo{volume}{108} (\bibinfo{year}{2015}) \bibinfo{pages}{1--5},
  \doi{\bibinfo{doi}{10.1016/j.scriptamat.2015.07.021}}.

\bibitem[{Curtarolo et~al.(2012{\natexlab{a}})Curtarolo, Setyawan, Hart,
  Jahn\'{a}tek, Chepulskii, Taylor, Wang, Xue, Yang, Levy, Mehl, Stokes,
  Demchenko, and Morgan}]{curtarolo:art65}
\bibinfo{author}{S.~Curtarolo}, \bibinfo{author}{W.~Setyawan},
  \bibinfo{author}{G.~L.~W. Hart}, \bibinfo{author}{M.~Jahn\'{a}tek},
  \bibinfo{author}{R.~V. Chepulskii}, \bibinfo{author}{R.~H. Taylor},
  \bibinfo{author}{S.~Wang}, \bibinfo{author}{J.~Xue},
  \bibinfo{author}{K.~Yang}, \bibinfo{author}{O.~Levy}, \bibinfo{author}{M.~J.
  Mehl}, \bibinfo{author}{H.~T. Stokes}, \bibinfo{author}{D.~O. Demchenko},
  \bibinfo{author}{D.~Morgan}, \bibinfo{title}{{AFLOW}: An automatic framework
  for high-throughput materials discovery}, \bibinfo{journal}{Comp.\ Mat.\
  Sci.} \bibinfo{volume}{58} (\bibinfo{year}{2012}{\natexlab{a}})
  \bibinfo{pages}{218--226},
  \doi{\bibinfo{doi}{10.1016/j.commatsci.2012.02.005}}.

\bibitem[{Jahn\'{a}tek et~al.(2011)Jahn\'{a}tek, Levy, Hart, Nelson,
  Chepulskii, Xue, and Curtarolo}]{curtarolo:art67}
\bibinfo{author}{M.~Jahn\'{a}tek}, \bibinfo{author}{O.~Levy},
  \bibinfo{author}{G.~L.~W. Hart}, \bibinfo{author}{L.~J. Nelson},
  \bibinfo{author}{R.~V. Chepulskii}, \bibinfo{author}{J.~Xue},
  \bibinfo{author}{S.~Curtarolo}, \bibinfo{title}{Ordered phases in ruthenium
  binary alloys from high-throughput first-principles calculations},
  \bibinfo{journal}{Phys.\ Rev.\ B} \bibinfo{volume}{84} (\bibinfo{year}{2011})
  \bibinfo{pages}{214110}, \doi{\bibinfo{doi}{10.1103/PhysRevB.84.214110}}.

\bibitem[{Chaput et~al.(2011)Chaput, Togo, Tanaka, and Hug}]{Chaput_prb_2011}
\bibinfo{author}{L.~Chaput}, \bibinfo{author}{A.~Togo},
  \bibinfo{author}{I.~Tanaka}, \bibinfo{author}{G.~Hug},
  \bibinfo{title}{Phonon-phonon interactions in transition metals},
  \bibinfo{journal}{Phys.\ Rev.\ B} \bibinfo{volume}{84} (\bibinfo{year}{2011})
  \bibinfo{pages}{094302}, \doi{\bibinfo{doi}{10.1103/PhysRevB.84.094302}}.

\bibitem[{Tadano et~al.(2014)Tadano, Gohda, and Tsuneyuki}]{Tadano_jpcm_2014}
\bibinfo{author}{T.~Tadano}, \bibinfo{author}{Y.~Gohda},
  \bibinfo{author}{S.~Tsuneyuki}, \bibinfo{title}{Anharmonic force constants
  extracted from first-principles molecular dynamics: applications to heat
  transfer simulations}, \bibinfo{journal}{J.\ Phys.:\ Conden.\ Matt.}
  \bibinfo{volume}{26} (\bibinfo{year}{2014}) \bibinfo{pages}{225402},
  \doi{\bibinfo{doi}{10.1088/0953-8984/26/22/225402}}.

\bibitem[{Curtarolo et~al.(2012{\natexlab{b}})Curtarolo, Setyawan, Wang, Xue,
  Yang, Taylor, Nelson, Hart, Sanvito, {Buongiorno~Nardelli}, Mingo, and
  Levy}]{curtarolo:art75}
\bibinfo{author}{S.~Curtarolo}, \bibinfo{author}{W.~Setyawan},
  \bibinfo{author}{S.~Wang}, \bibinfo{author}{J.~Xue},
  \bibinfo{author}{K.~Yang}, \bibinfo{author}{R.~H. Taylor},
  \bibinfo{author}{L.~J. Nelson}, \bibinfo{author}{G.~L.~W. Hart},
  \bibinfo{author}{S.~Sanvito}, \bibinfo{author}{M.~{Buongiorno~Nardelli}},
  \bibinfo{author}{N.~Mingo}, \bibinfo{author}{O.~Levy},
  \bibinfo{title}{AFLOWLIB.ORG: A distributed materials properties repository
  from high-throughput {\it ab initio} calculations}, \bibinfo{journal}{Comp.\
  Mat.\ Sci.} \bibinfo{volume}{58} (\bibinfo{year}{2012}{\natexlab{b}})
  \bibinfo{pages}{227--235},
  \doi{\bibinfo{doi}{10.1016/j.commatsci.2012.02.002}}.

\bibitem[{Taylor et~al.(2014)Taylor, Rose, Toher, Levy, Yang,
  {Buongiorno~Nardelli}, and Curtarolo}]{curtarolo:art92}
\bibinfo{author}{R.~H. Taylor}, \bibinfo{author}{F.~Rose},
  \bibinfo{author}{C.~Toher}, \bibinfo{author}{O.~Levy},
  \bibinfo{author}{K.~Yang}, \bibinfo{author}{M.~{Buongiorno~Nardelli}},
  \bibinfo{author}{S.~Curtarolo}, \bibinfo{title}{A {RESTful API} for
  exchanging Materials Data in the {AFLOWLIB.org} consortium},
  \bibinfo{journal}{Comp.\ Mat.\ Sci.} \bibinfo{volume}{93}
  (\bibinfo{year}{2014}) \bibinfo{pages}{178--192},
  \doi{\bibinfo{doi}{10.1016/j.commatsci.2014.05.014}}.

\bibitem[{Calderon et~al.(2015)Calderon, Plata, Toher, Oses, Levy, Fornari,
  Natan, Mehl, Hart, {Buongiorno~Nardelli}, and Curtarolo}]{curtarolo:art104}
\bibinfo{author}{C.~E. Calderon}, \bibinfo{author}{J.~J. Plata},
  \bibinfo{author}{C.~Toher}, \bibinfo{author}{C.~Oses},
  \bibinfo{author}{O.~Levy}, \bibinfo{author}{M.~Fornari},
  \bibinfo{author}{A.~Natan}, \bibinfo{author}{M.~J. Mehl},
  \bibinfo{author}{G.~L.~W. Hart}, \bibinfo{author}{M.~{Buongiorno~Nardelli}},
  \bibinfo{author}{S.~Curtarolo}, \bibinfo{title}{The {AFLOW} Standard for
  High-Throughput Materials Science Calculations}, \bibinfo{journal}{Comp.\
  Mat.\ Sci.} \bibinfo{volume}{108 Part A} (\bibinfo{year}{2015})
  \bibinfo{pages}{233--238},
  \doi{\bibinfo{doi}{10.1016/j.commatsci.2015.07.019}}.

\bibitem[{Orlikowski et~al.(2006)Orlikowski, S\"oderlind, and
  Moriarty}]{Orlikowski_prb_2006}
\bibinfo{author}{D.~Orlikowski}, \bibinfo{author}{P.~S\"oderlind},
  \bibinfo{author}{J.~A. Moriarty}, \bibinfo{title}{First-principles
  thermoelasticity of transition metals at high pressure: Tantalum prototype in
  the quasiharmonic limit}, \bibinfo{journal}{Phys.\ Rev.\ B}
  \bibinfo{volume}{74} (\bibinfo{year}{2006}) \bibinfo{pages}{054109},
  \doi{\bibinfo{doi}{10.1103/PhysRevB.74.054109}}.

\bibitem[{Xiang et~al.(2010)Xiang, Xi, Bi, Xu, Geng, Cai, Jing, and
  Liu}]{Xiang_prb2010}
\bibinfo{author}{S.~Xiang}, \bibinfo{author}{F.~Xi}, \bibinfo{author}{Y.~Bi},
  \bibinfo{author}{J.~Xu}, \bibinfo{author}{H.~Geng}, \bibinfo{author}{L.~Cai},
  \bibinfo{author}{F.~Jing}, \bibinfo{author}{J.~Liu},
  \bibinfo{title}{\textit{Ab initio} thermodynamics beyond the quasiharmonic
  approximation: {W} as a prototype}, \bibinfo{journal}{Phys.\ Rev.\ B}
  \bibinfo{volume}{81} (\bibinfo{year}{2010}) \bibinfo{pages}{014301},
  \doi{\bibinfo{doi}{10.1103/PhysRevB.81.014301}}.

\bibitem[{Grabowski et~al.(2009)Grabowski, Ismer, Hickel, and
  Neugebauer}]{PhysRevB.79.134106}
\bibinfo{author}{B.~Grabowski}, \bibinfo{author}{L.~Ismer},
  \bibinfo{author}{T.~Hickel}, \bibinfo{author}{J.~Neugebauer},
  \bibinfo{title}{{\it Ab initio} up to the melting point: Anharmonicity and
  vacancies in aluminum}, \bibinfo{journal}{Phys.\ Rev.\ B}
  \bibinfo{volume}{79}~(\bibinfo{number}{13}) (\bibinfo{year}{2009})
  \bibinfo{pages}{134106}, \doi{\bibinfo{doi}{10.1103/PhysRevB.79.134106}}.

\bibitem[{Skelton et~al.(2014)Skelton, Parker, Togo, Tanaka, and
  Walsh}]{Skelton_prb_2014}
\bibinfo{author}{J.~M. Skelton}, \bibinfo{author}{S.~C. Parker},
  \bibinfo{author}{A.~Togo}, \bibinfo{author}{I.~Tanaka},
  \bibinfo{author}{A.~Walsh}, \bibinfo{title}{Thermal physics of the lead
  chalcogenides PbS, PbSe, and PbTe from first principles},
  \bibinfo{journal}{Phys.\ Rev.\ B} \bibinfo{volume}{89} (\bibinfo{year}{2014})
  \bibinfo{pages}{205203}, \doi{\bibinfo{doi}{10.1103/PhysRevB.89.205203}}.

\bibitem[{Iikubo et~al.(2010)Iikubo, Ohtani, and Hasebe}]{Iikubo_mtr_2010}
\bibinfo{author}{S.~Iikubo}, \bibinfo{author}{H.~Ohtani},
  \bibinfo{author}{M.~Hasebe}, \bibinfo{title}{First-Principles Calculations of
  the Specific Heats of Cubic Carbides and Nitrides}, \bibinfo{journal}{Mater.\
  Trans.} \bibinfo{volume}{51} (\bibinfo{year}{2010})
  \bibinfo{pages}{574--577}, \doi{\bibinfo{doi}{10.2320/matertrans.MBW200913}}.

\bibitem[{de-la Roza and Lua{\~n}a(2011)}]{Roza_prb_2011}
\bibinfo{author}{A.~O. de-la Roza}, \bibinfo{author}{V.~Lua{\~n}a},
  \bibinfo{title}{Treatment of first-principles data for predictive
  quasiharmonic thermodynamics of solids: The case of MgO},
  \bibinfo{journal}{Phys.\ Rev.\ B} \bibinfo{volume}{84} (\bibinfo{year}{2011})
  \bibinfo{pages}{024109}, \doi{\bibinfo{doi}{10.1103/PhysRevB.84.024109}}.

\bibitem[{Tohei et~al.(2015)Tohei, Lee, and Ikuhara}]{Tohei_mtr_2015}
\bibinfo{author}{T.~Tohei}, \bibinfo{author}{H.-S. Lee},
  \bibinfo{author}{Y.~Ikuhara}, \bibinfo{title}{First Principles Calculation of
  Thermal Expansion of Carbon and Boron Nitrides Based on Quasi-Harmonic
  Approximation}, \bibinfo{journal}{Mater.\ Trans.} \bibinfo{volume}{56}
  (\bibinfo{year}{2015}) \bibinfo{pages}{1452--1456},
  \doi{\bibinfo{doi}{10.2320/matertrans.MA201574}}.

\bibitem[{Kangarlou and Abdollahi(2014)}]{Kangarlou_ijt_2014}
\bibinfo{author}{H.~Kangarlou}, \bibinfo{author}{A.~Abdollahi},
  \bibinfo{title}{Thermodynamic Properties of Copper in a Wide Range of
  Pressure and Temperature Within the Quasi-Harmonic Approximation},
  \bibinfo{journal}{Int.\ J.\ Thermophys.} \bibinfo{volume}{35}
  (\bibinfo{year}{2014}) \bibinfo{pages}{1501--1511},
  \doi{\bibinfo{doi}{10.1007/s10765-014-1742-x}}.

\bibitem[{Burton et~al.(2011)Burton, Demers, and van~de
  Walle}]{burton_jap_2011}
\bibinfo{author}{B.~P. Burton}, \bibinfo{author}{S.~Demers},
  \bibinfo{author}{A.~van~de Walle}, \bibinfo{title}{First principles phase
  diagram calculations for the wurtzite-structure quasibinary systems SiC-AlN,
  SiC-GaN and SiC-InN}, \bibinfo{journal}{J.\ Appl.\ Phys.}
  \bibinfo{volume}{110} (\bibinfo{year}{2011}) \bibinfo{pages}{023507},
  \doi{\bibinfo{doi}{10.1063/1.3602149}}.

\bibitem[{Golumbfskie et~al.(2006)Golumbfskie, Arroyave, Shin, and
  Liu}]{Golumbfskie_actamat_2006}
\bibinfo{author}{W.~J. Golumbfskie}, \bibinfo{author}{R.~Arroyave},
  \bibinfo{author}{D.~Shin}, \bibinfo{author}{Z.~K. Liu},
  \bibinfo{title}{Finite-temperature thermodynamic and vibrational properties
  of Al–Ni–Y compounds via first-principles calculations},
  \bibinfo{journal}{Acta\ Mater.} \bibinfo{volume}{54} (\bibinfo{year}{2006})
  \bibinfo{pages}{2291--2304},
  \doi{\bibinfo{doi}{http://dx.doi.org/10.1016/j.actamat.2006.01.013}}.

\bibitem[{Wee et~al.(2012)Wee, Kozinsky, Pavan, and Fornari}]{Wee_jem_2012}
\bibinfo{author}{D.~Wee}, \bibinfo{author}{B.~Kozinsky},
  \bibinfo{author}{B.~Pavan}, \bibinfo{author}{M.~Fornari},
  \bibinfo{title}{Quasiharmonic Vibrational Properties of TiNiSn from Ab Initio
  Phonons}, \bibinfo{journal}{J.\ Elec.\ Mat.} \bibinfo{volume}{41}
  (\bibinfo{year}{2012}) \bibinfo{pages}{977},
  \doi{\bibinfo{doi}{10.1007/s11664-011-1833-4}}.

\bibitem[{Agapito et~al.(2015)Agapito, Curtarolo, and
  {Buongiorno~Nardelli}}]{curtarolo:art93}
\bibinfo{author}{L.~A. Agapito}, \bibinfo{author}{S.~Curtarolo},
  \bibinfo{author}{M.~{Buongiorno~Nardelli}}, \bibinfo{title}{Reformulation of
  $\mathrm{DFT}+U$ as a Pseudohybrid Hubbard Density Functional for Accelerated
  Materials Discovery}, \bibinfo{journal}{Phys.\ Rev.\ X} \bibinfo{volume}{5}
  (\bibinfo{year}{2015}) \bibinfo{pages}{011006},
  \doi{\bibinfo{doi}{10.1103/PhysRevX.5.011006}}.

\bibitem[{Giannozzi et~al.(2009)Giannozzi, Baroni, Bonini, Calandra, Car,
  Cavazzoni, Ceresoli, Chiarotti, Cococcioni, Dabo, {Dal Corso}, {de
  Gironcoli}, Fabris, Fratesi, Gebauer, Gerstmann, Gougoussis, Kokalj, Lazzeri,
  Martin-Samos, Marzari, Mauri, Mazzarello, Paolini, Pasquarello, Paulatto,
  Sbraccia, Scandolo, Sclauzero, Seitsonen, Smogunov, Umari, and
  Wentzcovitch}]{quantum_espresso_2009}
\bibinfo{author}{P.~Giannozzi}, \bibinfo{author}{S.~Baroni},
  \bibinfo{author}{N.~Bonini}, \bibinfo{author}{M.~Calandra},
  \bibinfo{author}{R.~Car}, \bibinfo{author}{C.~Cavazzoni},
  \bibinfo{author}{D.~Ceresoli}, \bibinfo{author}{G.~L. Chiarotti},
  \bibinfo{author}{M.~Cococcioni}, \bibinfo{author}{I.~Dabo},
  \bibinfo{author}{A.~{Dal Corso}}, \bibinfo{author}{S.~{de Gironcoli}},
  \bibinfo{author}{S.~Fabris}, \bibinfo{author}{G.~Fratesi},
  \bibinfo{author}{R.~Gebauer}, \bibinfo{author}{U.~Gerstmann},
  \bibinfo{author}{C.~Gougoussis}, \bibinfo{author}{A.~Kokalj},
  \bibinfo{author}{M.~Lazzeri}, \bibinfo{author}{L.~Martin-Samos},
  \bibinfo{author}{N.~Marzari}, \bibinfo{author}{F.~Mauri},
  \bibinfo{author}{R.~Mazzarello}, \bibinfo{author}{S.~Paolini},
  \bibinfo{author}{A.~Pasquarello}, \bibinfo{author}{L.~Paulatto},
  \bibinfo{author}{C.~Sbraccia}, \bibinfo{author}{S.~Scandolo},
  \bibinfo{author}{G.~Sclauzero}, \bibinfo{author}{A.~P. Seitsonen},
  \bibinfo{author}{A.~Smogunov}, \bibinfo{author}{P.~Umari},
  \bibinfo{author}{R.~M. Wentzcovitch}, \bibinfo{title}{QUANTUM ESPRESSO: a
  modular and open-source software project for quantum simulations of
  materials}, \bibinfo{journal}{J.\ Phys.:\ Conden.\ Matt.}
  \bibinfo{volume}{21}~(\bibinfo{number}{39}) (\bibinfo{year}{2009})
  \bibinfo{pages}{395502}.

\bibitem[{Kresse and Hafner(1993)}]{kresse_vasp}
\bibinfo{author}{G.~Kresse}, \bibinfo{author}{J.~Hafner}, \bibinfo{title}{{\it
  Ab initio} molecular dynamics for liquid metals}, \bibinfo{journal}{Phys.\
  Rev.\ B} \bibinfo{volume}{47} (\bibinfo{year}{1993})
  \bibinfo{pages}{558--561}.

\bibitem[{Wallace(1972)}]{ThermoCrys}
\bibinfo{author}{D.~C. Wallace}, \bibinfo{title}{Thermodynamics of crystals},
  \bibinfo{publisher}{Wiley}, \bibinfo{year}{1972}.

\bibitem[{Srivastava(1990)}]{PhysPhon}
\bibinfo{author}{G.~P. Srivastava}, \bibinfo{title}{The Physics of Phonons},
  \bibinfo{publisher}{CRC Press, Taylor \& Francis}, \bibinfo{year}{1990}.

\bibitem[{Dove(1993)}]{IntroLattDym}
\bibinfo{author}{M.~T. Dove}, \bibinfo{title}{Introduction to Lattice
  Dynamics}, \bibinfo{publisher}{Cambridge University Press},
  \bibinfo{year}{1993}.

\bibitem[{Bl\"ochl(1994)}]{PAW}
\bibinfo{author}{P.~E. Bl\"ochl}, \bibinfo{title}{Projector augmented-wave
  method}, \bibinfo{journal}{Phys.\ Rev.\ B} \bibinfo{volume}{50}
  (\bibinfo{year}{1994}) \bibinfo{pages}{17953--17979}.

\bibitem[{Perdew et~al.(1996)Perdew, Burke, and Ernzerhof}]{PBE}
\bibinfo{author}{J.~P. Perdew}, \bibinfo{author}{K.~Burke},
  \bibinfo{author}{M.~Ernzerhof}, \bibinfo{title}{Generalized gradient
  approximation made simple}, \bibinfo{journal}{Phys.\ Rev.\ Lett.}
  \bibinfo{volume}{77} (\bibinfo{year}{1996}) \bibinfo{pages}{3865--3868}.

\bibitem[{Wang et~al.(2010)Wang, Wang, Wang, Mei, Shang, Chen, and
  Liu}]{Wang2010}
\bibinfo{author}{Y.~Wang}, \bibinfo{author}{J.~J. Wang}, \bibinfo{author}{W.~Y.
  Wang}, \bibinfo{author}{Z.~G. Mei}, \bibinfo{author}{S.~L. Shang},
  \bibinfo{author}{L.~Q. Chen}, \bibinfo{author}{Z.~K. Liu}, \bibinfo{title}{A
  mixed-space approach to first-principles calculations of phonon frequencies
  for polar materials}, \bibinfo{journal}{J. Phys.: Condens. Matter}
  \bibinfo{volume}{22}~(\bibinfo{number}{20}) (\bibinfo{year}{2010})
  \bibinfo{pages}{202201}, \doi{\bibinfo{doi}{10.1088/0953-8984/22/20/202201}}.

\bibitem[{B.Gauster(1971)}]{Gauster_prb_1971}
\bibinfo{author}{W.~B.Gauster}, \bibinfo{title}{Low-Temperature {G}r\"uneisen
  Parameters for Silicon and Aluminum}, \bibinfo{journal}{Phys.\ Rev.\ B}
  \bibinfo{volume}{4} (\bibinfo{year}{1971}) \bibinfo{pages}{1288--1296},
  \doi{\bibinfo{doi}{10.1103/PhysRevB.4.1288}}.

\bibitem[{Hughes and Cain(1996)}]{Hughes_prb_1996}
\bibinfo{author}{W.~C. Hughes}, \bibinfo{author}{L.~S. Cain},
  \bibinfo{title}{Second-order elastic constants of AgCl from 20 to
  430$^{\circ}$C}, \bibinfo{journal}{Phys.\ Rev.\ B} \bibinfo{volume}{53}
  (\bibinfo{year}{1996}) \bibinfo{pages}{5174},
  \doi{\bibinfo{doi}{http://dx.doi.org/10.1103/PhysRevB.53.5174}}.

\bibitem[{Barin(2008)}]{IhsanBarin}
\bibinfo{author}{I.~Barin}, \bibinfo{title}{Thermochemical Data of Pure
  Substances}, \bibinfo{publisher}{WILEY-VCH}, \bibinfo{year}{2008}.

\bibitem[{Madelung(2004)}]{Madelung_Semiconductors_2004}
\bibinfo{author}{O.~Madelung}, \bibinfo{title}{Semiconductors: Data Handbook},
  \bibinfo{publisher}{Springer Berlin Heidelberg}, \bibinfo{address}{Berlin},
  \bibinfo{year}{2004}.

\bibitem[{Slack(1979)}]{slack}
\bibinfo{author}{G.~A. Slack}, \bibinfo{title}{The thermal conductivity of
  nonmetallic crystals}, in: \bibinfo{editor}{H.~Ehrenreich},
  \bibinfo{editor}{F.~Seitz}, \bibinfo{editor}{D.~Turnbull} (Eds.),
  \bibinfo{booktitle}{Solid State Physics}, vol.~\bibinfo{volume}{34},
  \bibinfo{publisher}{Academic, New York}, \bibinfo{pages}{1},
  \bibinfo{year}{1979}.

\bibitem[{Lide(2004)}]{Lide_CRC_2004}
\bibinfo{author}{D.~R. Lide}, \bibinfo{title}{CRC Handbook of Chemistry and
  Physics}, \bibinfo{publisher}{Taylor \& Francis}, \bibinfo{year}{2004}.

\bibitem[{Laplaze et~al.(1976)Laplaze, Boissier, and Vacher}]{Laplaze_ssc_1976}
\bibinfo{author}{D.~Laplaze}, \bibinfo{author}{M.~Boissier},
  \bibinfo{author}{R.~Vacher}, \bibinfo{title}{Velocity of hypersounds in
  lithium hydride by spontaneous Brillouin scattering}, \bibinfo{journal}{Solid
  State Commun.} \bibinfo{volume}{19} (\bibinfo{year}{1976})
  \bibinfo{pages}{445--446}, \doi{\bibinfo{doi}{10.1016/0038-1098(76)91187-X}}.

\bibitem[{Lam et~al.(1987)Lam, Cohen, and Martinez}]{Lam_prb_1987}
\bibinfo{author}{P.~K. Lam}, \bibinfo{author}{M.~L. Cohen},
  \bibinfo{author}{G.~Martinez}, \bibinfo{title}{Analytic relation between bulk
  moduli and lattice constants}, \bibinfo{journal}{Phys.\ Rev.\ B}
  \bibinfo{volume}{35} (\bibinfo{year}{1987}) \bibinfo{pages}{9190},
  \doi{\bibinfo{doi}{http://dx.doi.org/10.1103/PhysRevB.35.9190}}.

\bibitem[{McNeil et~al.(1993)McNeil, Grimsditch, and
  French}]{McNeil_jacers_1993}
\bibinfo{author}{L.~E. McNeil}, \bibinfo{author}{M.~Grimsditch},
  \bibinfo{author}{R.~H. French}, \bibinfo{title}{Vibrational Spectroscopy of
  Aluminum Nitride}, \bibinfo{journal}{J.\ Am.\ Ceramic.\ Soc.}
  \bibinfo{volume}{76} (\bibinfo{year}{1993}) \bibinfo{pages}{1132--1136},
  \doi{\bibinfo{doi}{10.1111/j.1151-2916.1993.tb03730.x}}.

\bibitem[{Morelli and Slack(2006)}]{Morelli_Slack_2006}
\bibinfo{author}{D.~T. Morelli}, \bibinfo{author}{G.~A. Slack},
  \bibinfo{title}{High Lattice Thermal Conductivity Solids}, in:
  \bibinfo{editor}{S.~L. Shind{\'e}}, \bibinfo{editor}{J.~S. Goela} (Eds.),
  \bibinfo{booktitle}{High Thermal Conductivity Materials},
  \bibinfo{publisher}{Springer}, \bibinfo{year}{2006}.

\bibitem[{Hauss{\"u}hl(1960)}]{Hauss_zfp_1960}
\bibinfo{author}{S.~Hauss{\"u}hl}, \bibinfo{title}{Thermo-elastische Konstanten
  der Alkalihalogenide vom NaCl-Typ}, \bibinfo{journal}{Z.\ f{\"u}r \ Physik}
  \bibinfo{volume}{159} (\bibinfo{year}{1960}) \bibinfo{pages}{223--229},
  \doi{\bibinfo{doi}{10.1007/BF01338349}}.

\bibitem[{Li(1976)}]{Li_jpcrd_1976}
\bibinfo{author}{H.~H. Li}, \bibinfo{title}{Refractive Index of alkali halides
  and its wavelength and temperature derivatives}, \bibinfo{journal}{J.\ Phys.\
  Chem.\ Ref. \ Data} \bibinfo{volume}{5} (\bibinfo{year}{1976})
  \bibinfo{pages}{329--528}.

\bibitem[{Ueno et~al.(1994)Ueno, Yoshida, Onodera, Shimomura, and
  Takemura}]{Ueno_prb_1994}
\bibinfo{author}{M.~Ueno}, \bibinfo{author}{M.~Yoshida},
  \bibinfo{author}{A.~Onodera}, \bibinfo{author}{O.~Shimomura},
  \bibinfo{author}{K.~Takemura}, \bibinfo{title}{Stability of the wurtzite-type
  structure under high pressure: GaN and InN}, \bibinfo{journal}{Phys.\ Rev.\
  B} \bibinfo{volume}{49} (\bibinfo{year}{1994}) \bibinfo{pages}{14--21},
  \doi{\bibinfo{doi}{http://dx.doi.org/10.1103/PhysRevB.49.14}}.

\bibitem[{Krukowski et~al.(1998)Krukowski, Witek, Adamczyk, Jun, Bockowski,
  Grzegory, Lucznik, Nowak, Wr{\'o}blewski, Presz, Gierlotka, Stelmach, Palosz,
  Porowski, and Zinn}]{Krukowski_jphyschemsolids_1998}
\bibinfo{author}{S.~Krukowski}, \bibinfo{author}{A.~Witek},
  \bibinfo{author}{J.~Adamczyk}, \bibinfo{author}{J.~Jun},
  \bibinfo{author}{M.~Bockowski}, \bibinfo{author}{I.~Grzegory},
  \bibinfo{author}{B.~Lucznik}, \bibinfo{author}{G.~Nowak},
  \bibinfo{author}{M.~Wr{\'o}blewski}, \bibinfo{author}{A.~Presz},
  \bibinfo{author}{S.~Gierlotka}, \bibinfo{author}{S.~Stelmach},
  \bibinfo{author}{B.~Palosz}, \bibinfo{author}{S.~Porowski},
  \bibinfo{author}{P.~Zinn}, \bibinfo{title}{Thermal properties of indium
  nitride}, \bibinfo{journal}{J.\ Phys.\ Chem.\ Solids}
  \bibinfo{volume}{59}~(\bibinfo{number}{3}) (\bibinfo{year}{1998})
  \bibinfo{pages}{289--295},
  \doi{\bibinfo{doi}{10.1016/S0022-3697(97)00222-9}}.

\bibitem[{Xu et~al.(2013)Xu, Wang, and Tian}]{Xu_sr_2013}
\bibinfo{author}{B.~Xu}, \bibinfo{author}{Q.~Wang}, \bibinfo{author}{Y.~Tian},
  \bibinfo{title}{Bulk modulus for polar covalent crystals},
  \bibinfo{journal}{Sci.\ Rep.} \bibinfo{volume}{3} (\bibinfo{year}{2013})
  \bibinfo{pages}{3068}, \doi{\bibinfo{doi}{10.1038/srep03068}}.

\bibitem[{Sumino et~al.(1976)Sumino, Ohno, Goto, and
  Kumazawa}]{Sumino_jpearth_1976}
\bibinfo{author}{Y.~Sumino}, \bibinfo{author}{I.~Ohno},
  \bibinfo{author}{T.~Goto}, \bibinfo{author}{M.~Kumazawa},
  \bibinfo{title}{Measurement of elastic constants and internal frictions on
  single-crystal MgO by rectangular parallelepiped resonance},
  \bibinfo{journal}{J.\ Phys.\ Earth} \bibinfo{volume}{24}
  (\bibinfo{year}{1976}) \bibinfo{pages}{263--273},
  \doi{\bibinfo{doi}{http://doi.org/10.4294/jpe1952.24.263}}.

\bibitem[{Chang and Graham(1977)}]{Chang_jphyschemsolids_1977}
\bibinfo{author}{Z.~P. Chang}, \bibinfo{author}{E.~K. Graham},
  \bibinfo{title}{Elastic properties of oxides in the NaCl-structure},
  \bibinfo{journal}{J.\ Phys.\ Chem.\ Solids} \bibinfo{volume}{38}
  (\bibinfo{year}{1977}) \bibinfo{pages}{1355--1362},
  \doi{\bibinfo{doi}{10.1016/0022-3697(77)90007-5}}.

\bibitem[{Carrete et~al.(2014)Carrete, Li, Mingo, Wang, and
  Curtarolo}]{curtarolo:art84}
\bibinfo{author}{J.~Carrete}, \bibinfo{author}{W.~Li},
  \bibinfo{author}{N.~Mingo}, \bibinfo{author}{S.~Wang},
  \bibinfo{author}{S.~Curtarolo}, \bibinfo{title}{Finding Unprecedentedly
  Low-Thermal-Conductivity Half-Heusler Semiconductors via High-Throughput
  Materials Modeling}, \bibinfo{journal}{Phys.\ Rev.\ X} \bibinfo{volume}{4}
  (\bibinfo{year}{2014}) \bibinfo{pages}{011019},
  \doi{\bibinfo{doi}{10.1103/PhysRevX.4.011019}}.

\bibitem[{Gheribi et~al.(2015)Gheribi, Seifitokaldani, Wu, and
  Chartrand}]{Gheribi_jap_2015}
\bibinfo{author}{A.~E. Gheribi}, \bibinfo{author}{A.~Seifitokaldani},
  \bibinfo{author}{P.~Wu}, \bibinfo{author}{P.~Chartrand}, \bibinfo{title}{An
  ab initio method for the prediction of the lattice thermal transport
  properties of oxide systems: Case study of {Li$_2$O} and {K$_2$O}},
  \bibinfo{journal}{J.\ Appl.\ Phys.} \bibinfo{volume}{118}
  (\bibinfo{year}{2015}) \bibinfo{pages}{145101},
  \doi{\bibinfo{doi}{10.1063/1.4932643}}.

\bibitem[{Orabi et~al.(2016)Orabi, Mecholsky, Hwang, Kim, Rhyee, Wee, and
  Fornari}]{Orabi_cm_2016}
\bibinfo{author}{R.~A. R.~A. Orabi}, \bibinfo{author}{N.~A. Mecholsky},
  \bibinfo{author}{J.~Hwang}, \bibinfo{author}{W.~Kim}, \bibinfo{author}{J.-S.
  Rhyee}, \bibinfo{author}{D.~Wee}, \bibinfo{author}{M.~Fornari},
  \bibinfo{title}{Band Degeneracy, Low Thermal Conductivity, and High
  Thermoelectric Figure of Merit in {S}n{T}e-{C}a{T}e},
  \bibinfo{journal}{Chem.\ Mat.} \bibinfo{volume}{28} (\bibinfo{year}{2016})
  \bibinfo{pages}{376--384}, \doi{\bibinfo{doi}{DOI:
  10.1021/acs.chemmater.5b04365}}.

\bibitem[{Vaqueiro et~al.(2015)Vaqueiro, Orabi, Luu, Gu\'elou, Powell, Smith,
  Song, Wee, and Fornari}]{Vaqueiro_pccp_2015}
\bibinfo{author}{P.~Vaqueiro}, \bibinfo{author}{R.~A. R.~A. Orabi},
  \bibinfo{author}{S.~D.~N. Luu}, \bibinfo{author}{G.~Gu\'elou},
  \bibinfo{author}{A.~V. Powell}, \bibinfo{author}{R.~I. Smith},
  \bibinfo{author}{J.-P. Song}, \bibinfo{author}{D.~Wee},
  \bibinfo{author}{M.~Fornari}, \bibinfo{title}{The role of copper in the
  thermal conductivity of thermoelectric oxychalcogenides: do lone pairs
  matter?}, \bibinfo{journal}{Phys.\ Chem.\ Chem.\ Phys.} \bibinfo{volume}{17}
  (\bibinfo{year}{2015}) \bibinfo{pages}{31735},
  \doi{\bibinfo{doi}{10.1039/C5CP06192J}}.

\bibitem[{Srivastava et~al.(2009)Srivastava, Sinha, and
  Panwar}]{SKSrivastavaIJoPaAP}
\bibinfo{author}{S.~K. Srivastava}, \bibinfo{author}{P.~Sinha},
  \bibinfo{author}{M.~Panwar}, \bibinfo{title}{Thermal expansivity and
  isothermal bulk modulus of ionic materials at high temperatures},
  \bibinfo{journal}{Indian J. Pure Appl. Phys.} \bibinfo{volume}{47}
  (\bibinfo{year}{2009}) \bibinfo{pages}{175--179}.

\bibitem[{Hermet et~al.(2014)Hermet, Ayral, Theron, Yot, Salles, Tillard, and
  Jund}]{PHermetJPhysChem}
\bibinfo{author}{P.~Hermet}, \bibinfo{author}{R.~M. Ayral},
  \bibinfo{author}{E.~Theron}, \bibinfo{author}{P.~G. Yot},
  \bibinfo{author}{F.~Salles}, \bibinfo{author}{M.~Tillard},
  \bibinfo{author}{P.~Jund}, \bibinfo{title}{Thermal Expansion of {Ni-Ti-Sn}
  Heusler and Half-Heusler Materials from First-Principles Calculations and
  Experiments}, \bibinfo{journal}{J. Phys. Chem. C}
  \bibinfo{volume}{118}~(\bibinfo{number}{39}) (\bibinfo{year}{2014})
  \bibinfo{pages}{22405--22411}, \doi{\bibinfo{doi}{10.1021/jp502112f}}.

\bibitem[{Blakemore(1982)}]{Blakemore_jap1982}
\bibinfo{author}{J.~S. Blakemore}, \bibinfo{title}{Semiconducting and other
  major properties of gallium arsenide}, \bibinfo{journal}{J.\ Appl.\ Phys.}
  \bibinfo{volume}{53} (\bibinfo{year}{1982}) \bibinfo{pages}{R123--R181},
  \doi{\bibinfo{doi}{10.1063/1.331665}}.

\end{thebibliography}
\newcommand{\Ozolins}{Ozoli\c{n}\v{s}}

\end{document}


\begin{frontmatter}
  \title{\large Supplementary Materials: High-Throughput Prediction of Finite-Temperature Properties using the Quasi-Harmonic Approximation}
  \author{Pinku Nath$^1$, Jose J. Plata$^1$, Demet Usanmaz$^1$, Rabih Al Rahal Al Orabi$^2$, Marco Fornari$^2$, Marco Buongiorno Nardelli$^3$, Cormac Toher$^1$, Stefano Curtarolo$^{7,\star}$}
  \address{$^{1}$ Department of Mechanical Engineering and Materials Science, Duke University, Durham, North Carolina 27708, USA.}
  \address{$^{2}$ Department of Physics and Science of Advanced Materials Program, Central Michigan University, Mount Pleasant, MI 48858, USA.}
  \address{$^{3}$ Department of Physics and Department of Chemistry, University of North Texas, Denton TX, USA.}
  \address{$^{4}$ Materials Science, Electrical Engineering, Physics and Chemistry, Duke University, Durham NC, 27708, USA.}
  \address{$^{\star}${\bf corresponding:} stefano@duke.edu}
\end{frontmatter}

\renewcommand{\theequation}{S\arabic{equation}}
\renewcommand{\thesection} {S\Roman{section}}
\renewcommand{\thefigure}  {S\arabic{figure}}
\renewcommand{\thetable}   {S\arabic{table}}
\renewcommand{\bibnumfmt}[1]{[S#1]}
\renewcommand{\citenumfont}[1]{S#1}

\section*{Band gaps}\label{introduction}

\begin{longtable}[!p]{c r l l | c r l l }
  \caption{\small ICSD identification number, theoretical and experimental band gap values, E$^{GGA}_{gap}$ and E$^{exp}_{gap}$ for material data set. Units: E$^{GGA}_{gap}$ and E$^{exp}_{gap}$ in eV.}\\
\hline  Formula           &  ICSD    &    E$^{GGA}_{gap}$ & E$^{exp}_{gap}$  &  Formula         &  ICSD      & E$^{GGA}_{gap}$      & E$^{exp}_{gap}$  \\ \hline
\endfirsthead
\hline  Formula           &  ICSD    &    E$^{GGA}_{gap}$ & E$^{exp}_{gap}$  &  Formula         &  ICSD      & E$^{GGA}_{gap}$      & E$^{exp}_{gap}$  \\ \hline
\endhead
\hline
\endfoot
\hline \hline
\endlastfoot
  \ce{Ag1Cl1}      & 157535   &  2.01             &   3.0\cite{Duffy_ssc_1984}                                            &
  \ce{Ge1}         &  44841   &  0.07             &   0.7\cite{Nethercot_prl_1974}                                        \\
  \ce{Ag1Mg1}      & 184205   &  Metal            &   Metal                                                                  &
  \ce{Ge1Mg2}      &  81735   &  0.23             &   0.74\cite{Lide_CRC_2004}                                            \\
  \ce{Ag1Sc1}      & 58348    &  Metal            &   Metal                                                                 &
  \ce{H1Li1}       &  61751   &  3.09             &   4.9\cite{Plekhanov_ssc_1990}                                        \\
  \ce{Ag3Mg1}      & 58323    &  Metal            &   Metal                                                                  &
  \ce{H1Li1Pd1}    &  246613  &  Metal            &   --                                                                  \\
  \ce{Al1As1}      & 606008   &  1.54             &   2.16\cite{Safa_KasapHandbookofElectronic}                           &
  \ce{H1Mg1Ni1}    &  187257  &  Metal            &   --                                                                  \\
  \ce{Al1B2}       & 159334   &  Metal            &   --                                                               &
  \ce{H1Na1}       &  183291  &  3.82             &   --                                                                  \\
  \ce{Al1}         & 240129   &  Metal            &   Metal                                                                  &
  \ce{H1Ti1}       &  168325  &  Metal            &   Metal                                                                  \\
  \ce{Al1Li1}      & 240121   &  Metal            &   Metal                                                                  &
  \ce{Hg1Ni1}      &  639119  &  Metal            &   Metal                                                                  \\
  \ce{Al1N1}       & 602459   &  4.0              &   --                                                                  &
  \ce{Hg1Pd1}      &  639137  &  Metal            &   Metal                                                                  \\
  \ce{Al1Ni1}      & 608805   &  Metal            &   Metal                                                               &
  \ce{Hg1Pt1}      &  104337  &  Metal            &   Metal                                                                  \\
  \ce{Al1P1}       & 609019   &  1.67             &   3.63\cite{Safa_KasapHandbookofElectronic}                           &
  \ce{Hg1Zr1}      &  639318  &  Metal            &   Metal                                                                  \\
  \ce{Al1Sb1}      & 609290   &  1.24             &   1.5\cite{Duffy_ssc_1984}                                            &
  \ce{I1K1}        &  53827   &  3.86             &   6.0\cite{Nethercot_prl_1974}                                          \\
  \ce{Al1Sc1}      & 58098    &  Metal            &   Metal                                                                   &
  \ce{I1Li1}       &  27983   &  4.26             &   5.8\cite{Nethercot_prl_1974}                                        \\
  \ce{Al1Si1Sr1}   & 162865   &  Metal            &   --                                                                      &
  \ce{I1Na1}       &  52240   &  3.64             &   5.6\cite{Duffy_ssc_1984}                                             \\
  \ce{Al1Tb3}      & 58173    &  Metal            &   Metal                                                                   &
  \ce{I1Rb1}       &  53846   &  3.79             &   5.8\cite{Duffy_ssc_1984}                                             \\
  \ce{Al1Ti1}      & 187030   &  Metal            &   Metal                                                                   &
  \ce{In1N1}       &  157515  &  Metal            &  --                                                                    \\
  \ce{Al3Ti1}      & 609525   &  Metal            &   Metal                                                                   &
  \ce{In1P1}       &  165466  &  0.73             &   1.30\cite{Nethercot_prl_1974}                                        \\
  \ce{As1B1}       & 181292   &  1.27             &   --                                                                   &
  \ce{In1Te1}      &  169431  &  Metal            &   --                                                                   \\
  \ce{As1Ba1Li1}   & 56445    &  0.59             &   --                                                                   &
  \ce{In1Te1}      &  640622  &  Metal            &   --                                                                   \\
  \ce{As1Ga1}      & 53964    &  0.57             &   1.35\cite{Lide_CRC_2004}                                             &
  \ce{K1}          &  641218  &  Metal            &   Metal                                                                   \\
  \ce{As1In1}      & 165462   &  0.19             &   0.38\cite{Nethercot_prl_1974}                                        &
  \ce{K2O1}        &  44674   &  1.71             &   --                                                                   \\
  \ce{B1Sb1}       & 184571   &  0.81             &   --                                                                   &
  \ce{K2S1}        &  183837  &  2.32             &   --                                                                   \\
  \ce{B2Ti1}       & 78847    &  Metal            &   Metal                                                                   &
  \ce{Li1Pd1}      &  642257  &  Metal            &   Metal                                                                  \\
  \ce{B2V1}        & 167794   &  Metal            &   Metal                                                                  &
  \ce{Li1Pt1}      &  104777  &  Metal            &   Metal                                                               \\
  \ce{Be1}         & 52708    &  Metal            &   Metal                                                                  &
  \ce{Li2O1}       &  60431   &  4.93             &   --                                                                  \\
  \ce{Be1Rh1}      & 58734    &  Metal            &   Metal                                                                  &
  \ce{Li2S1}       &  657596  &  3.41             &   --                                                                  \\
  \ce{Be1S1}       & 186889   &  3.19             &   --                                                                  &
  \ce{Li2Se1}      &  168446  &  3.00             &   --                                                                  \\
  \ce{Be1Se1}      & 616419   &  2.71             &  5.6\cite{Madelung_Semiconductors_2004}                               &
  \ce{Li2Te1}      &  642399  &  2.52             &  --                                                                   \\
  \ce{Be1Te1}      & 290008   &  2.06             &  --                                                                   &
  \ce{Mg1O1}       &  159372  &  4.53             &  7.83\cite{Aryasetiawan_PRB_1995}                                     \\
  \ce{Be2C1}       & 616184   &  1.21             &  --                                                                   &
  \ce{Mg1Pt3}      &  104857  &  Metal            &  Metal                                                                \\
  \ce{Bi1Na1}      & 616837   &  Metal            &  --                                                                    &
  \ce{Mg1S1}       &  53939   &  2.80             &  --                                                                   \\
  \ce{Br1Cu1}      & 30090    &  1.19             &  3.0\cite{Duffy_ssc_1984}                                             &
  \ce{Mg1Sc1}      &  108583  &  Metal            &  Metal                                                                   \\
  \ce{Br1K1}       & 52243    &  4.34             &  7.8\cite{Nethercot_prl_1974}                                         &
  \ce{Mg1Se1}      &  159398  &  Metal            &  --                                                                      \\
  \ce{Br1Li1}      & 53819    &  4.95             &  7.2 \cite{Duffy_ssc_1984}                                            &
  \ce{Mg2Pb1}      &  104846  &  Metal            &  Metal                                                                   \\
  \ce{Br1Na1}      & 44278    &  4.14             &  6.7\cite{Duffy_ssc_1984}                                             &
  \ce{Mg2Si1}      &  163708  &  0.26             &  0.78\cite{Morris_pr_1958}                                               \\
  \ce{C1}          & 182729   &  4.22             &  5.4\cite{Lide_CRC_2004}                                              &
  \ce{Mg2Sn1}      &  151368  &  Metal            &  Metal                                                                   \\
  \ce{C1Si1}       & 618777   &  2.36             &  2.9\cite{Duffy_ssc_1984}                                             &
  \ce{Mn1O1}       &  18006   &  0.25             &  3.6-3.8\cite{Svane_prl_1990}                                         \\
  \ce{C1Ti1}       & 181681   &  Metal            &  --                                                                    &
  \ce{Mn1S1}       &  76205   &  1.35             &  3.5\cite{Tsai_prb_1996}                                              \\
  \ce{C1Zr1}       & 180599   &  Metal            &  --                                                                     &
  \ce{Mn1Se1}      &  24252   &  0.93             &  2.0-2.5\cite{Youn_pssb_2004}                                         \\
  \ce{Ca1Cd1}      & 619188   &  Metal            &  Metal                                                                &
  \ce{Mn1Te1}      &  181324  &  1.13             &  0.9-1.3\cite{Youn_pssb_2004}                                         \\
  \ce{Ca1F2}       & 40938    &  7.12             &  10.0\cite{Smith_prb_1977}                                            &
  \ce{N1Sc1}       &  155049  &  0.54             &  --                                                                   \\
  \ce{Ca1O1}       & 180198   &  3.67             &  7.0\cite{Whited_ssc_1973}                                             &
  \ce{N1Ti1}       &  183415  &  Metal            &  Metal                                                                   \\
  \ce{Ca1S1}       & 619530   &  2.40             &  --                                                                   &
  \ce{Na1}         &  644903  &  Metal            &  Metal                                                                   \\
  \ce{Ca1Se1}      & 619570   &  2.10             &  --                                                                   &
  \ce{Se1Zn1}      &  181761  &  1.79             &  2.58\cite{Lide_CRC_2004}                                             \\
  \ce{Ca1Te1}      & 619616   &  1.58             &  --                                                                   &
  \ce{Ni1O1}       &  166115  &  1.82             &  --                                                                   \\
  \ce{Cd1F2}       & 28864    &  3.05             &  8.4\cite{Cappellini_prb_2013}                                        &
  \ce{Ni1Sb1}      &  646431  &  Metal            &  Metal                                                                   \\
  \ce{Cd1O1}       & 181735   &  Metal            &  --                                                                   &
  \ce{Ni1Sc1}      &  105333  &  Metal            &  Metal                                                                   \\
  \ce{Cd1Pd1}      & 620270   &  Metal            &  Metal                                                                   &
  \ce{Ni1Zn1}      &  647134  &  Metal            &  Metal                                                                \\
  \ce{Cd1Pt1}      & 620297   &  Metal            &  Metal                                                                   &
  \ce{Ni2Sn1Ti1}   &  646777  &  Metal            &  --                                                                     \\
  \ce{Cd1S1}       & 290009   &  1.26             &  2.6\cite{Duffy_ssc_1984}                                             &
  \ce{O1Pd1}       &  26598   &  0.22             &  1.5\cite{Rogers_jssc_1971}                                           \\
  \ce{Cd1Sr1}      & 102066   &  Metal            &  Metal                                                                   &
  \ce{O1Sr1}       &  26960   &  3.30             &  5.9\cite{Taurian_ssc_1985}                                                            \\
  \ce{Cd3Zr1}      & 102093   &  Metal            &  Metal                                                                &
  \ce{O1Zn1}       &  182356  &  1.81             &  3.2\cite{Lide_CRC_2004}                                              \\
  \ce{Cl1Cu1}      & 78270    &  1.31             &  3.17\cite{Lide_CRC_2004}                                             &
  \ce{O1Zn1}       &  647683  &  1.76             &  3.4 \cite{Duffy_ssc_1984}                                            \\
  \ce{Cl1K1}       & 240522   &  5.06             &  7.8\cite{Duffy_ssc_1984}                                             &
  \ce{Pd1Zn1}      &  649134  &  Metal            &  Metal                                                                \\
  \ce{Cl1Li1}      & 26909    &  6.26             &  8.8\cite{Duffy_ssc_1984}                                             &
  \ce{Pt1Zn1}      &  105852  &  Metal            &  Metal                                                                   \\
  \ce{Cl1Na1}      & 240600   &  5.06             &  8.7\cite{Nethercot_prl_1974}                                         &
  \ce{S1Zn1}       &  108733  &  2.74             &  3.62\cite{Nethercot_prl_1974}                                        \\
  \ce{Cl1Rb1}      & 18016    &  4.85             &  7.5\cite{Duffy_ssc_1984}                                             &
  \ce{Sc1}         &  164093  &  Metal            &  Metal                                                                   \\
  \ce{Cu1I1}       & 163427   &  1.67             &  3.1 \cite{Safa_KasapHandbookofElectronic}                            &
  \ce{Sc1Zn1}      &  106041  &  Metal            &  Metal                                                                   \\
  \ce{Cu1}         & 627117   &  Metal            &  Metal                                                                   &
  \ce{Si1}         &  76268   &  0.66             &  1.10\cite{Nethercot_prl_1974}                                        \\
  \ce{Cu1Sn1}      & 629278   &  Metal            &  Metal                                                                   &
  \ce{Tc1V1}       &  106143  &  Metal            &  Metal                                                                   \\
  \ce{Cu2Ni1Zn1}   & 103079   &  Metal            &  --                                                                     &
  \ce{Te1Zn1}      &  184485  &  1.56             &  2.26\cite{LevIBerger_SemiconductorMaterials}                         \\
  \ce{F1K1}        & 52241    &  5.98             &  11.0\cite{Nethercot_prl_1974}                                        &
  \ce{Te1Zr1}      &  653209  &  Metal            &  --                                                                   \\
  \ce{F1Li1}       & 53839    &  8.75             &  12.9\cite{Duffy_ssc_1984}                                            &
  \ce{Ti1}         &  168830  &  Metal            &  Metal                                                                   \\
  \ce{F1Na1}       & 52238    &  6.15             &  10.8\cite{Duffy_ssc_1984}                                            &
  \ce{Zn1}         &  181734  &  Metal            &  Metal                                                                   \\
  \ce{Ga1P1}       & 77088    &  1.66             &  2.2\cite{Duffy_ssc_1984}                                             &
  \ce{Zn1Zr1}      &  181290  &  Metal            &  Metal                                                                   \\
  \ce{Ga1Sb1}      & 41675    &  0.30             &  0.70\cite{Nethercot_prl_1974}                                        &
                   &          &                   &  --                                                                   \\
\label{tab:data}
\end{longtable}

\bibliographystyle{elsarticle-num-names}
\newcommand{\Ozolins}{Ozoli\c{n}\v{s}}